\documentclass[lettersize,journal]{IEEEtran}
\usepackage{amsmath,amsfonts,amssymb}
\usepackage{algorithmic}
\usepackage{algorithm}
\usepackage{array}
\usepackage[caption=false,font=normalsize,labelfont=sf,textfont=sf]{subfig}
\usepackage{textcomp}
\usepackage{stfloats}
\usepackage{url}
\usepackage{verbatim}
\usepackage{graphicx}
\usepackage{cite}
\usepackage{float}
\usepackage[table]{xcolor}
\usepackage{amsthm}
\usepackage{ascmac}
\usepackage{enumitem}
\usepackage{tabularx}
\usepackage{multirow}
\usepackage{etoolbox}
\usepackage{lipsum}
\usepackage{flushend}
    
\newtheorem{Theorem}{Theorem}
\newtheorem{Definition}{Definition}
\newtheorem{Remark}{Remark}
\newtheorem{Lemma}{Lemma}

\newtheorem{Corollary}{Corollary}
\newtheorem{Proposition}{Proposition}

\begin{document}

\title{Outer Bounds on the CEO Problem with \\ Privacy Constraints}

\author{Vamoua~Yachongka,~\IEEEmembership{Member,~IEEE,} Hideki~Yagi,~\IEEEmembership{Member,~IEEE,} Hideki~Ochiai,~\IEEEmembership{Fellow,~IEEE}
\thanks{V.\ Yachongka is with the Department of Computer Science and Engineering, The University of Texas at Arlington, Arlington, TX 76019 USA. Corresponding author: Vamoua Yachongka (e-mail: va.yachongka@ieee.org).}
\thanks{H. Yagi is with the Department of Computer and Network Engineering, The University of Electro-Communications, Chofu, Tokyo, 182-8585 Japan. Email: h.yagi@uec.ac.jp}
\thanks{H.\ Ochiai is affiliated with Graduate School of Engineering, Osaka University, 2-1 Yamadaoka, Suita, Osaka, 565-0871, Japan. Email: ochiai@comm.eng.osaka-u.ac.jp}
}



\maketitle

\begin{abstract}
We investigate the rate-distortion-leakage region of the Chief Executive Officer (CEO) problem, considering the presence of a passive eavesdropper and privacy constraints. We start by examining the region where a general distortion measure quantifies the distortion. While the inner bound of the region is derived from previous work, this paper newly develops an outer bound. To derive the outer bound, we introduce a new lemma tailored for analyzing privacy constraints. Next, as a specific instance of the general distortion measure, we demonstrate that the tight bound for discrete and Gaussian sources is obtained when the eavesdropper has no side information, and the distortion is quantified by the log-loss distortion measure. We further investigate the rate-distortion-leakage region for a scenario where the eavesdropper has side information, and the distortion is quantified by the log-loss distortion measure and provide an outer bound for this case. The derived outer bound differs from the inner bound by only a minor quantity that appears in the constraints associated with the privacy-leakage rates, and these bounds match when the distortion is large.
\end{abstract}

\begin{IEEEkeywords}
Eavesdropper, logarithmic loss, privacy leakage, rate region, the CEO problem.
\end{IEEEkeywords}

\section{Introduction}
\begin{figure}[!t]
\centering
   \includegraphics[scale=0.45]{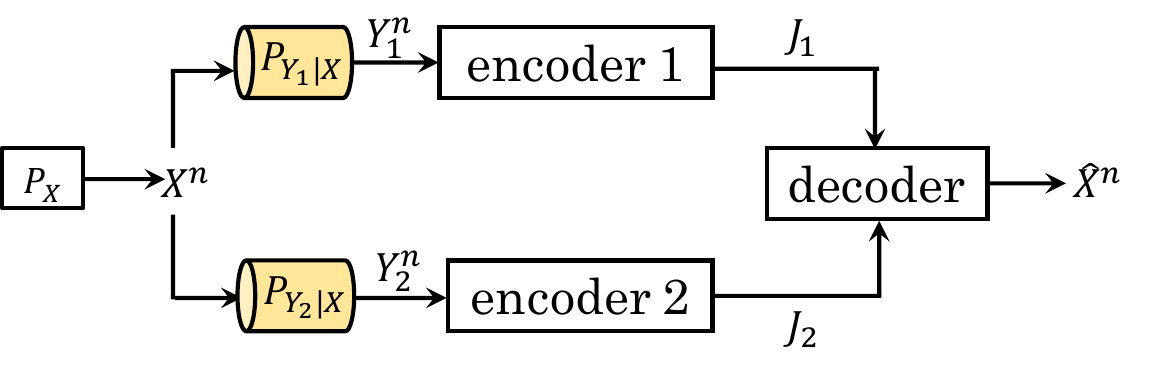}
   \vspace{-4mm}
   \caption{The CEO problem with two encoders where $Y^n_1$ and $Y^n_2$ denote measurements of source sequence $X^n$ observed through channels $P_{Y_1|X}$ and $P_{Y_2|X}$, respectively. The compressed messages $J_1$ and $J_2$ are transmitted to the central decoder over noiseless channels. The decoder utilizes these messages to reconstruct $\hat{X}^n$, a reproduced sequence of the source $X^n$.}
   \label{CEO-model}
   \vspace{-4mm}
\end{figure}

\IEEEPARstart{T}{he} Chief Executive Officer (CEO) problem, introduced by Berger et al. in \cite{berger1996}, can be viewed as a particular case of multiterminal source coding \cite{Berger1978, tung1978}. As depicted in Fig. \ref{CEO-model}, multiple encoders observe noisy measurements of a remote source and transmit compressed messages separately to the decoder. The task of the decoder is to reconstruct the original source sequence within a prescribed distortion level, taking into account the independent noises added during the observation process. Practical applications related to this problem encompass a broad range of areas, with a focus on distributed sensor networks \cite{kong2010} and collaborative communications \cite{lin2021}. For example, photos of an individual captured by infrared sensor security cameras are compressed in a distributed manner, and the compressed sequences are sent parallelly to a control unit for reconstructing the individual’s image.
\subsection{Related Works}
In the early 1980s, an achievable rate-distortion region of the CEO problem with two encoders was studied in \cite{yamamoto1980}. Later, an alternative expression of the achievable region and some code-design techniques were developed in \cite{FlynnGray1978}. The problem was generalized by Berger et al. in \cite{berger1996} for discrete sources. The case of Gaussian sources and quadratic distortion was studied in \cite{viswanathan}, where the asymptotic trade-off between the sum rate and the distortion was characterized. This characterization was subsequently refined in \cite{oohama1998}. The CEO problem has been broadly investigated in many subsequent papers, but even to date, finding a tight rate-distortion region for a general distortion measure remains to be solved.

For some special cases, however, notable progress has been made. The tight rate-distortion region for the quadratic Gaussian CEO problem has been elegantly delineated in \cite{Chen2004,Oohama2005,Prab2004}. Numerous extensions of the scalar Gaussian case are found in \cite{wagner2008,byzantineCEO,oohama2012,oohama2014,chenetal2017,chen2020,kistina2021,bilen2023}. The vector Gaussian source, for which the tight rate-distortion region has not been clarified, is extensively discussed in \cite{chen2011,wang2012,Ekrem2014,wang2014-tit}. Furthermore, some new results have been presented for the non-regular CEO problem in \cite{vempaty2015}, as well as for the regular and non-regular CEO problems incorporating a generalized distortion metric in \cite{seo2021-tit}.
Logarithmic loss (log-loss) distortion, which allows the decoder to reproduce the sequence generated by the remote source with a soft decision, emerges as an alternative distortion metric that yields the tight bound for the CEO problem \cite{courtade2014}. Based on this fundamental result, a practical coding scheme is proposed in \cite{nagnir2018,nangir2019-tcom} for the binary CEO problem. Expansions to continuous alphabets for complete characterizations of the logarithmic Gaussian CEO problem for scalar and vector Gaussian sources can be found in \cite{seo2016} and \cite{ugur2020}, respectively.   

In physical layer security, privacy protection of sensitive information sources is important and commonly arises in many real-life applications such as secure packet transmissions in distributed sensor networks as well as smart grid systems \cite{shankar2013}, privacy in databases \cite{sankar-tifs-2013,shinihara2023}, confidential image sharing in satellite communications \cite{aljohani2016}, and key agreement using physical unclonable functions (PUFs) \cite{gunlu-tifs-2018,vyo2023-tifs}. See also \cite{tyagi2012-jsac,tu2019-tit,chou2022,gunlu-tit2022} for discussions of multi-party function computations with privacy. Taking privacy issues into account, multiterminal source coding in the presence of a passive eavesdropper (Eve) has been investigated in \cite{vilard2013,naghibis2015}, where Eve can obtain not only the messages sent over the noise-free link connected among encoders and the decoder but also a correlated sequence of the source, called side information (SI). In \cite{naghibis2015}, compared to the conventional setups where the compression-rate and distortion constraints are imposed, secrecy (equivocation) constraints are newly added in the problem formulation of the CEO problem to evaluate the maximum equivocation rates between the source sequence and the information available at Eve. The authors in \cite{naghibis2015} derive inner and outer bounds on the rate-distortion-equivocation region under a general distortion measure, and as an example, they also derive an achievable region for the model where the messages and SI are available at Eve, as well as the tight region for a simpler model where Eve does not have access to the correlated sequence under the quadratic distortion.

\subsection{Motivations}

The model discussed in this paper aligns with the one described in \cite{naghibis2015}, which addresses the two-link CEO problem involving an adversary, Eve. In this scenario, Eve intercepts communications from the encoder to the decoder and has access to a sequence that is correlated with the hidden source. A practical application of this model can be seen in cooperative communication, where two drones capture photos of a sensitive object, such as a military vehicle, and wirelessly transmit encoded messages to a control unit that aims to reconstruct a photo of the object. Due to the broadcast nature of wireless transmissions, Eve might intercept these transmissions and may also have a noisy image of the object, possibly an old picture of the object taken in the past. Using the intercepted message and the noisy image, Eve attempts to gain knowledge of the sensitive object. The key objective is to minimize the compression rates required for the drones to transmit the compressed data to the central unit, while also minimizing the amount of privacy leakage about the sensitive object to Eve.

Our model is the same as the one considered in \cite{naghibis2015}, but the constraints imposed in the definition of problem formulation differ. Specifically, instead of equivocation constraints, we impose privacy constraints, evaluating the minimal amount of leakage on the measurement observed at the encoder gained by Eve in our problem formulation of the CEO problem. The rationale for replacing equivocation constraints is that when log-loss distortion is considered, the tight bound on the rate-distortion-equivocation region may not be achievable even for the model without SI at Eve (refer to Appendix \ref{appendixE} for a counterexample). On the other hand, privacy constraints have been successfully analyzed in the context of multiterminal source coding with an untrusted helper \cite{kitti2016}, where the tight bound was attained under log-loss distortion. This result hints at the possibility of deriving a tight bound for the CEO problem by replacing the equivocation constraints in \cite{naghibis2015} with privacy constraints, which motivates our approach.

We note that the difference between privacy and equivocation constraints is a conditional entropy. By a simple deduction of the inner bound and outer bounds of the rate-distortion-equivocation region under a general distortion measure derived in \cite[Theorems 1 and 2]{naghibis2015}, we can obtain the counterparts of the rate region for the CEO problem with privacy constraints, called the rate-distortion-leakage region\footnote{The rate-distortion-leakage region differs from the rate-distortion-equivocation region by incorporating privacy constraints in the achievability definition (Definition 1) of the CEO problem instead of equivocation constraints.} in this paper. However, the resulting expression for the outer bound is somehow simplistic and offers limited insights into the possibility of deriving a tighter bound for the log-loss case. For this reason, we are interested in exploring a novel outer bound on the rate-distortion-leakage region for a general distortion measure.

As a special case of a general distortion, we study the rate-distortion-leakage region under the log-loss distortion measure. Although the log-loss distortion measure falls into a special instance, its derivation is not obtained from a trivial reduction of the general case and instead requires an independent approach \cite{courtade2014}. In the context of the CEO problem without privacy constraints, a complete characterization of the tight bound was possible in several setups \cite{nagnir2018,seo2016,ugur2020}. However, when privacy constraints are taken into account, it remains unknown whether the tight bound for the CEO problem can be derived, which drives our focus on investigating the rate-distortion-leakage region under this criterion.



\subsection{Main Contributions}

The main contributions of this paper are summarized as follows:
\begin{itemize}
    \item We derive a novel outer bound on the rate-distortion-leakage region of the CEO problem when the distortion is quantified by a general distortion measure. Compared to the outer bound obtained by a simple reduction from \cite[Theorem 2]{naghibis2015}, this new bound bears a closer resemblance to the expression of the inner bound. To characterize the outer bound, we establish a lemma to bound the privacy-leakage rates and their sum rate explicitly.
    \item We demonstrate that it is possible to characterize a tight bound for the model without SI at Eve under log-loss distortion for both discrete and Gaussian sources by evaluating the privacy-leakage rates. Moreover, numerical examples are conducted for a case where compression rates are fixed. The obtained results reveal that the distortion is in a saturation state as the privacy-leakage rate becomes large, implying that further improvement on the distortion is not possible. In addition, as the distortion decreases, the compression rates as well as the privacy-leakage rate increase.
    \item We provide an extended version of the lemma used for analyzing the outer bound under a general distortion measure and apply it to characterize an outer bound under the log-loss distortion for the model with SI at Eve. The derived bound does not generally match the inner bound, but the difference between these bounds is only a quantity that appears in the constraints of the privacy-leakage rates and their sum rate. When the distortion is large, the inner and outer bounds become~tight.
\end{itemize}

For notation, uppercase $X$ and lowercase $x$ denote a random variable and its realization, respectively. The probability mass function of a random variable $X$ is written as $P_{X}(\cdot)$. The superscript $n$ denotes $n$-length sequences across time, e.g., $X^n = (X_1,X_2,\cdots,X_n)$, where the subscript represents the position of a random variable in the string. Denote $X^{n\backslash t} = (X_1,\cdots,X_{t-1},X_{t+1},\cdots,X_n)$ and $X_{\mathcal{A}}=(X_k : k\in \mathcal{A})$ for a set $\mathcal{A}$.
For integers $j$ and $k$ such that (s.t.) $j<k$, $[j:k]=\{j,j+1,\cdots,k\}$. The closed interval between real numbers $x$ and $y$ is denoted by $[x,y]$. The set of nonnegative real numbers is denoted as $\mathbb{R}_+$, $\mathbb{E}[\cdot]$ denotes the expected value of a random variable, and $\log x~(x > 0)$ is the base of two.

The rest of this paper is organized as follows: We describe our system model and formulate the problem definition in Section \ref{sect2}. An outer bound on the rate-distortion-leakage region under a general distortion measure is given in Section \ref{sect3}. Section \ref{sect4} presents the statement of our main results under log-loss distortion measure, and finally, concluding remarks and some future directions are given in Section \ref{sect5}.

\section{System Model and Problem Formulation} \label{sect2}
\begin{figure*}[!t]
\centering
   \begin{minipage}{0.49\textwidth}
   \includegraphics[width=\textwidth]{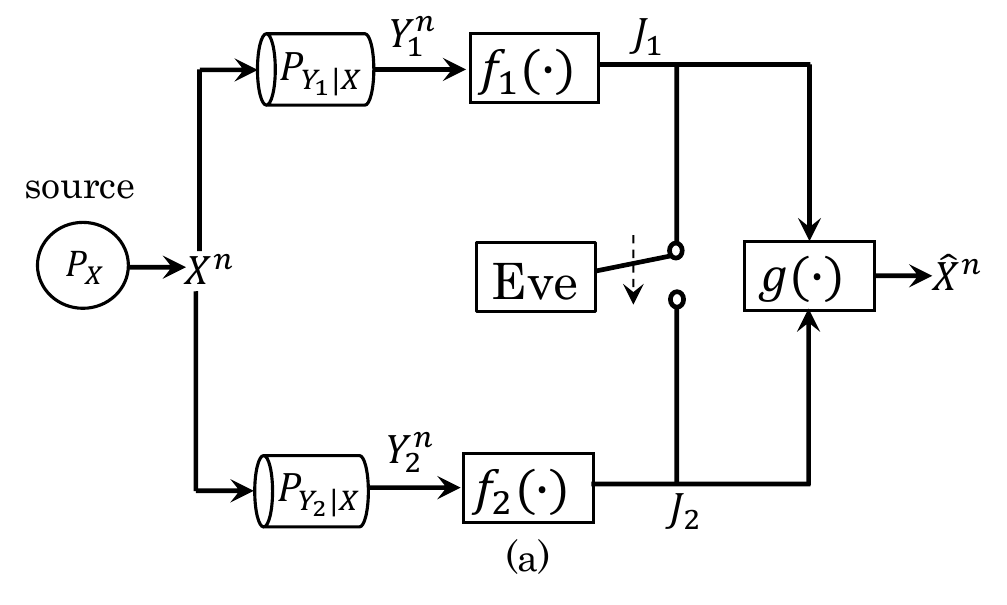}
   \end{minipage}
   \begin{minipage}{0.49\textwidth}
   \includegraphics[width=\textwidth]{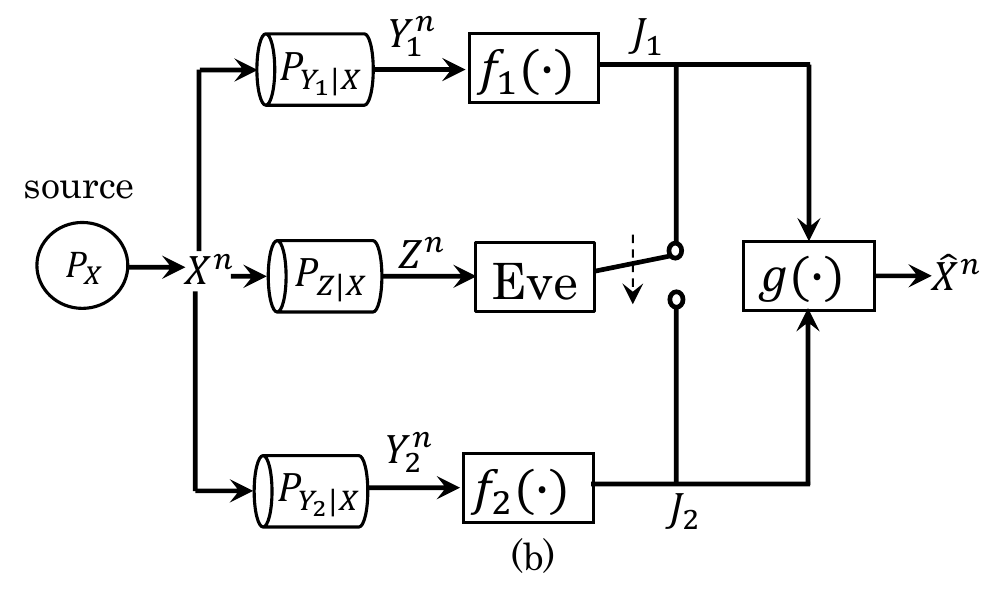}
   \end{minipage}
   \vspace{-3mm}
   \caption{System model of the CEO problem in the presence of a passive eavesdropper; (a) when Eve has no side information of the hidden source and (b) when Eve has side information of the hidden source.}
   \label{model}
\end{figure*}

\subsection{System Model}
Our system models are described in Figs.\ \ref{model}(a) and \ref{model}(b). The former figure corresponds to the model where SI is not available at Eve, i.e., the channel to Eve does not exist, and the latter one corresponds to the model where SI is available at Eve. Thus, the difference between the two models is the data Eve has.
Random vectors $X^n$ and $\hat{X}^n$ stand for the source sequence and its reconstructed version, respectively. Assume that the sequence $X^n$ is independently and identically distributed (i.i.d.) generated from the source $P_X$, and the channel to the first encoder (agent) $f_1$ $(\mathcal{X},P_{Y_1|X}, \mathcal{Y}_1)$, the channel to the second agent $f_2$ $(\mathcal{X},P_{Y_2|X}, \mathcal{Y}_2)$, and the channel to Eve $(\mathcal{X},P_{Z|X},\mathcal{Z})$ are all discrete and memoryless. The joint distributions of models in Figs. \ref{model}(a) and \ref{model}(b) are given by
\vspace{-2mm}
\begin{align}
    P_{X^nY^n_1Y^n_2} 
    &= \textstyle \prod_{t=1}^n P_{Y_{1,t}|X_t}\cdot P_{X_t} \cdot P_{Y_{2,t}|X_t}, \label{jointd-no-SI} \\
    P_{X^nY^n_1Y^n_2Z^n} 
    &= \textstyle \prod_{t=1}^n P_{Y_{1,t}|X_t}\cdot P_{X_t} \cdot P_{Y_{2,t}|X_t}\cdot P_{Z_t|X_t}. \label{jointd}
\end{align}
For $k \in \{1,2\}$, let $\mathcal{J}_k$ be the set of messages generated by agent $f_k$. We basically assume that all alphabets $\mathcal{Y}_k$, $\mathcal{X}$, and $\mathcal{Z}$ are finite, which will be relaxed for the derivation of Gaussian sources in Section \ref{tight-bound-gs}.

The agent $f_k$ generates the message $J_k \in \mathcal{J}_k$, which we denote as
$
    J_k = f_k(Y^n_k).
$
The message $J_k$ can be regarded as compressed versions of $Y^n_k$, and both $J_1$ {and} $J_2$ are passed to the decoder $g$ (the CEO or the fusion center) through public noiseless channels, {over} which the sent information can be completely intercepted by Eve. Upon receiving the messages, the CEO reconstructs a lossy version of the original source sequence as
$
    \hat{X}^n = g(J_1,J_2).
$

The distortion measure between the source and reconstructed sequences is expressed as
\vspace{-3mm}
\begin{align}
    d^{(n)}(x^n,\hat{x}^n) = \frac{1}{n}\sum_{t=1}^nd(x_t,\hat{x}_t), \label{dis-equation}
\end{align}
where $d:\mathcal{X}\times \hat{\mathcal{X}} \rightarrow \mathbb{R}_+$ is {a} single-letter measure.

In the model depicted in Fig. \ref{model}(a), Eve passively eavesdrops on $J_1$ and $J_2$, while in Fig. \ref{model}(b), Eve observes $Z^n$, called side information (SI) in this paper, and eavesdrops $J_1$ and $J_2$. In each scenario, it is assumed that Eve can only observe one of $J_1$ or $J_2$ at a time, and she leverages the observed message (along with the SI $Z^n$ for model depicted in Fig. \ref{model}(b)) to infer the source sequence $X^n$, which remains hidden from public networks. In this paper, our objective is to evaluate the minimum amount of leakage of $X^n$ to Eve.
\vspace{-2mm}
\subsection{Problem Formulation} \label{II-B}
The achievability definition of our model is defined as follows:
\vspace{-2mm}
\begin{Definition} \label{defsystem}
A tuple of compression rates, privacy-leakage rates, and distortion $(R_1,R_2,L_1,L_2,D) \in \mathbb{R}^5_+$ is said to be achievable if for $k \in \{1,2\}$, $\delta > 0$, and large enough $n$, there exist encoders $(f_1,f_2)$ and decoder $g$ satisfying
\vspace{-2mm}
\begin{align}
   \log|\mathcal{J}_k| &\leq n(R_k + \delta), \label{storage} \hspace{5mm} {\rm ({compression})}\\
   I(X^n;J_k|Z^n) &\leq n({L_k+\delta}), \label{pl} \hspace{5.5mm} {\rm (privacy~leakage)}\\
   \textstyle \mathbb{E}[{d^{(n)}}(X^n,\hat{X}^n)] &\leq D + \delta, \label{distortion} \hspace{12mm} {\rm (distortion)}
\end{align}
\qed
\end{Definition}
The technical meaning of each constraint in Definition \ref{defsystem} can be interpreted as follows: The constraint \eqref{storage} {is the compression constraint}, and it is imposed to minimize the set of messages to reduce the transmission and hardware cost; \eqref{pl} {is the privacy constraint and} evaluates the minimal amount of privacy leakage of the source sequence to Eve\footnote{In Definition \ref{defsystem}, \eqref{pl} assesses the privacy leakage at each agent separately. The analysis of joint leakage, represented by $I(X^n;J_1,J_2|Z^n)$, will be a subject of future research.}; and \eqref{distortion} {is the distortion constraint and} indicates that the distortion between $X^n$ and $\hat{X}^n$ should not exceed $D$.

For $k \in \{1,2\}$, the notations $R_k$, $L_k$, and $D$ represent the minimum asymptotic values that quantify the right-hand sides of constraints \eqref{storage}, \eqref{pl}, and \eqref{distortion}, respectively, when the block length is sufficiently large. These values are referred to in this paper as the compression rate, the privacy-leakage rate, and the distortion, respectively.

\begin{Remark} For the model in Fig. \ref{model}(a), the problem definition satisfies all conditions in Definition \ref{defsystem} with \eqref{pl} replaced by
\begin{align}
    I(X^n;J_k) \leq n({L_k+\delta}). \label{pl-no-SI}
\end{align}
\end{Remark}
\begin{Definition} \label{region-r}
$\mathcal{R}$ is defined as the closure of the set of all achievable rate tuples, called the rate-distortion-leakage region in this paper. Moreover, inner and outer bounds of $\mathcal{R}$ are denoted as $\mathcal{R}_{\rm in}$ and $\mathcal{R}_{\rm out}$, respectively, where it holds that $\mathcal{R}_{\rm in} \subseteq \mathcal{R}$ and $\mathcal{R}_{\rm out} \supseteq \mathcal{R}$. For a special case where the log-loss distortion is used, we deploy $\mathcal{R}^{\rm L}_{\rm in}$ and $\mathcal{R}^{\rm L}_{\rm out}$ to denote the inner and outer bounds, respectively. \qed
\end{Definition}

In this paper, we explore the region $\mathcal{R}$ and its inner and outer bounds for different scenarios on model settings, source types, and distortion measures. To aid in understanding each scenario, we outline the following cases and subcases.
\begin{itemize}
    \item[(I)] When the distortion is measured using a general distortion measure, we examine inner and outer bounds on the region $\mathcal{R}$ of the model with SI at Eve (Fig. \ref{model}(b)) for discrete sources. This case is discussed in Section \ref{sect3}.
    \item[(II)] When the distortion is quantified by the log-loss distortion measure,
    \begin{itemize}
        \item[(II.1)] the region $\mathcal{R}$ of the model without SI at Eve (Fig. \ref{model}(a)) is derived for discrete sources, addressed in Section~\ref{tight-bound-ds}.
        \item[(II.2)] the region $\mathcal{R}$ of the model without SI at Eve (Fig. \ref{model}(a)) is derived for Gaussian sources, addressed in Section~\ref{tight-bound-gs}.
        \item[(II.3)] inner and outer bounds on the region $\mathcal{R}$ of the model with SI at Eve (Fig. \ref{model}(b)) for discrete sources are derived, addressed in Section \ref{si-at-eve}.
    \end{itemize}
\end{itemize}
We will again mention these cases in the relevant sections and subsections where they apply.
\vspace{-2mm}
\subsection{{Relation to Previous Works}}
\begin{table}[t!]
    {\caption{\label{table-ii-c} Comparison of the constraints in the problem definitions between some highly relevant previous studies and our work}
    \vspace{-2mm}
        \begin{tabular}{| m{1cm} | m{0.6cm} m{1.7mm}  m{1.7mm}  m{1.7mm} m{1.1cm}m{1.1cm}m{1.1cm}|}
            \hline
             Paper & Model & $R_k$& $\Delta_k$& $L_k$& general distortion &quadratic distortion & log-loss distortion\\ 
            \hline
            \cite{Berger1978},\cite{tung1978} & (a) & \checkmark & &  & \checkmark & & \\
            \hline
        \vspace{-3mm}\cite{courtade2014},~\cite{seo2016} & (a) & \checkmark & &  & & & \checkmark\\ 
            \hline
            \multirow{2}*{\cite{naghibis2015}} & (a) & \checkmark & \checkmark & & & \checkmark~ & \\ 
            \cline{2-8}
             & (b) & \checkmark & \checkmark & & \checkmark & \checkmark & \\
            \hline
            \multirow{2}*{Our work} & (a) & \checkmark & & \checkmark & & & \checkmark\\
            \cline{2-8}
            & (b) & \checkmark & & \checkmark & \checkmark & & \checkmark\\
            \hline
        \end{tabular}}
        \vspace{-3mm}
\end{table}
A comparison of constraints and distortion in the problem definitions between some highly relevant previous studies and our work is summarized in Table \ref{table-ii-c}. The problem definitions in \cite{Berger1978},\cite{tung1978},\cite{courtade2014},\cite{seo2016} are special cases of our problem. Later, in the presentation of our main results, we will clarify that the findings of these studies can be derived from our main result by disregarding the privacy constraints.

In \cite{naghibis2015}, the authors characterized an inner bound on the rate-distortion-equivocation region, obtained when \eqref{pl} is replaced by
\vspace{-2mm}
\begin{align}
    H(X^n|J_k,Z^n) \geq n({\Delta_k-\delta}), \label{equi-no-SI}
\end{align}
where $\Delta_k$ denotes the equivocation rate for $k \in \{1,2\}$. The result of \cite[Theorem 1]{naghibis2015} serves as the foundation~for deriving inner bounds in our work, and thus we present~the detailed expression of it for reference. The time-sharing random variable $Q$, independent of $(X,Y_1,Y_2,Z)$, is used for convexification.
\begin{Theorem} \label{th-pw} {(\hspace{-0.1mm}\cite[Theorem 1]{naghibis2015}) An inner bound on the rate-distortion-equivocation region for a general distortion measure is given by the closure of the set of all rate tuples $(R_1,R_2,\Delta_1,\Delta_2,D)\in \mathbb{R}^5_+$ s.t.
\begin{align}
    R_1 &\ge I(Y_1;U_1|U_2,Q),~
    R_2 \ge I(Y_2;U_2|U_1,Q), \nonumber \\
    R_1 + R_2 &\ge I(Y_1,Y_2;U_1,U_2|Q),\nonumber \\
    \Delta_1 &\le H(X|Z) - I(X;U_1|U_2,Q) \nonumber \\
    &~~~~- I(V_1;U_2|Q) + I(Z;V_1|Q), \label{l1-cons-pw} \\
    \Delta_2 &\le H(X|Z) - I(X;U_2|U_1,Q) \nonumber \\
    &~~~~+ I(V_2;U_1|Q)-I(Z;V_2|Q), \label{l2-cons-pw} \\
    \Delta_1+\Delta_2 &\le 2H(X|Z) - I(X;U_1,U_2|Q) - I(V_1;V_2|Q) \nonumber \\
    &~~~~+I(Z;V_1|Q) +I(Z;V_2|Q), \label{l1l2-cons-pw} \\
    R_1 + \Delta_2 &\le H(X|Z) - I(Y_1,X;U_1,U_2|Q) \nonumber \\
    &~~~~+ I(Z;V_2|Q), \label{r1l2-cons-pw} \\
    R_2 + \Delta_1 & \le H(X|Z) - I(X,Y_2;U_1,U_2|Q) \nonumber \\
    &~~~~ + I(Z;V_1|Q), \label{r2l1-cons-pw} \\
    D &\ge \mathbb{E}[d(X,\hat{X}(U_1,U_2,Q))] \label{dis-cons-pw}
\end{align}
for all joint distributions $P_{V_1U_1Y_1XZY_2U_2V_2Q}$ factorized as
\vspace{-2mm}
\begin{align}
\textstyle P_{X}P_{Z|X}P_{Q}\prod_{{k}=1}^2P_{Y_k|X}P_{U_k|Y_kQ}P_{V_k|U_kQ} \label{jdist-inner}
\end{align}
with $|\mathcal{V}_k| \le |\mathcal{Y}_k| + 9$, $|\mathcal{U}_k| \le (|\mathcal{Y}_k| + 9)(|\mathcal{Y}_k| + 5)$, $|\mathcal{Q}| \le 6$, and a reproduction function $\hat{X}: \mathcal{U}_1\times \mathcal{U}_2 \times \mathcal{Q} \longrightarrow \hat{\mathcal{X}}$.}
\end{Theorem}
\begin{proof}
    The detailed proof is discussed in \cite[Appendix A]{naghibis2015}. For a detailed expansion of the right-hand side of each constraint in \cite[Theorem 1]{naghibis2015} with the Markov chains defined in \eqref{jdist-inner}, the readers are referred to \cite[Appendix G]{naghibis2015}. Here, we only mention the idea of the proof. The proof is based on layered coding, where the auxiliary sequences $V^n_k$ and $U^n_k$ generated by the agent $k$ are divided and saved in bins and sub-bins. The agent $k$ shares the bin and sub-bin indexes with the decoder via noise-free channels, so the decoder is able to reliably estimate $V^n_k$ and $U^n_k$ to reconstruct $\hat{X}^n$.
\end{proof}


\section{Inner and Outer Bounds for General Distortion Measure} \label{sect3}
In this section, we present inner and outer bounds on the rate-distortion-leakage region under a general distortion measure. The scenario considered here corresponds to Case (I) in Section \ref{II-B}, and the joint distribution of the model is given in \eqref{jointd}. The inner bound directly stems from the result presented in \cite[Theorem 1]{naghibis2015}. In contrast, the outer bound is newly derived, and its expression shares more similarity with the inner bound compared to the one obtained by a simple deduction from \cite[Theorem 2]{naghibis2015}.

The inner bound on the rate-distortion-leakage region is given below.
\begin{Theorem} \label{th1}
An inner bound $\mathcal{R}_{\rm in}$ on the rate-distortion-leakage region {for a general distortion measure} is characterized as the closure of the set of all rate tuples $(R_1,R_2,L_1,L_2,D)\in \mathbb{R}^5_+$ s.t.
\begin{align}
    R_1 &\ge I(Y_1;U_1|U_2,Q),~
    R_2 \ge I(Y_2;U_2|U_1,Q), \label{r2-cons} \\
    R_1 + R_2 &\ge I(Y_1,Y_2;U_1,U_2|Q),\label{r1r2-cons} \\
    L_1 &\ge I(X;U_1|U_2,Q) \nonumber \\
    &~~~~+ I(V_1;U_2|Q)-I(Z;V_1|Q), \label{l1-cons-1} \\
    L_2 &\ge I(X;U_2|U_1,Q) \nonumber \\
    &~~~~+ I(V_2;U_1|Q)-I(Z;V_2|Q), \label{l2-cons-2} \\
    L_1+L_2 &\ge I(X;U_1,U_2|Q) + I(V_1;V_2|Q) \nonumber \\
    &~~~~-I(Z;V_1|Q) -I(Z;V_2|Q), \label{l1l2-cons} \\
    R_1 + L_2 &\ge I(Y_1,X;U_1,U_2|Q) - I(Z;V_2|Q), \label{r1l2-cons} \\
    R_2 + L_1 & \ge I(X,Y_2;U_1,U_2|Q) - I(Z;V_1|Q), \label{r2l1-cons} \\
    D &\ge \mathbb{E}[d(X,\hat{X}(U_1,U_2,Q))], \label{dis-cons}
\end{align}
where the auxiliary random variables $(V_1,V_2,U_1,U_2)$, their respective cardinalities, and the reproduction function $\hat{X}(U_1, U_2, Q)$ satisfy the same conditions as in Theorem \ref{th-pw}.
\end{Theorem}
\begin{proof}
Theorems 1 and 2 have a similarity in terms of constraints on compression rates, sum rates, and distortion, allowing for similar derivations. However, they differ in the constraints related to privacy-leakage and equivocation rates. For details, see \eqref{l1-cons-pw}–\eqref{r2l1-cons-pw} in Theorem 1 and \eqref{l1-cons-1}–\eqref{r2l1-cons} in Theorem 2. Since the privacy constraint in \eqref{pl} and equivocation constraint in \eqref{equi-no-SI} are related by $\frac{1}{n}I(X^n;J_k|Z^n) = H(X|Z) - \frac{1}{n}H(X^n|J_k,Z^n)$, Theorem \ref{th1} is derived by subtracting the right-hand sides of \eqref{l1-cons-pw}, \eqref{l2-cons-pw}, \eqref{r1l2-cons-pw}, and \eqref{r2l1-cons-pw} in Theorem \ref{th-pw} from $H(X|Z)$ and \eqref{l1l2-cons-pw} in Theorem \ref{th-pw} from $2H(X|Z)$.
\end{proof}

The constraints \eqref{r2-cons}, \eqref{r1r2-cons}, and \eqref{dis-cons} in Theorem \ref{th1} coincide with the Berger-Tung inner bound \cite{Berger1978,tung1978} when the auxiliary sequences $(V^n_1,U^n_1,U^n_2,V^n_2)$ are reliably estimated at the decoder within a certain distortion level $D$. Equation \eqref{l2-cons-2} expresses the minimal amounts of the privacy-leakage rate at each agent, and \eqref{l1l2-cons} is their sum rate. Conditions \eqref{r1l2-cons} and \eqref{r2l1-cons} reflect trade-offs of the compression and privacy-leakage rates of different agents. This trade-off suggests that in terms of the sum rate, the inner region of the CEO problem with privacy constraints is not only constrained by the sum rate of compression rates and the sum rate of privacy-leakage rates, but also a jointly mixed combination of these rates from distinct agents.

\begin{table*}[t]
\centering
\caption{Comparison of Rate Constraints in Inner and Outer Bounds For General Distortion Measure}
\label{table-g}
\vspace{-2mm}
\begin{tabular}{| m{1cm} | m{7cm}| m{8.8cm} | } 
  \hline
    \hfil Rates& \hfil Inner bound & \hfil Outer bound \\ 
   \hline
  \rowcolor{gray!10}
   $R_1$ & $I(Y_1;U_1|U_2,Q)$ & $I(Y_1;U_1|U_2,Q)$ \\ 
  $R_2$ & $I(Y_2;U_2|U_1,Q)$ & $I(Y_2;U_2|U_1,Q)$ \\ 
  \rowcolor{gray!10}
  $R_1 + R_2$ & $I(Y_1,Y_2;U_1,U_2|Q)$ & $I(Y_1,Y_2;U_1,U_2|Q)$ \\
  $L_1$ & $I(X;U_1|U_2,Q) + I(V_1;U_2|Q)-I(Z;V_1|Q)$ & $I(X;V_1|V_2,Q) + I(V_1;U_2|Q)-I(Z;V_1|Q) - \xi_1$ \\
  \rowcolor{gray!10}
  $L_2$ & $I(X;U_2|U_1,Q) + I(V_2;U_1|Q)-I(Z;V_2|Q)$ & $I(X;V_2|V_1,Q) + I(V_2;U_1|Q)-I(Z;V_2|Q) - \xi_2$ \\
  $L_1 + L_2$ & $I(X;U_1,U_2|Q) + I(V_1;V_2|Q)-I(Z;V_1|Q) -I(Z;V_2|Q)$ & $I(X;V_1,V_2|Q) + I(V_1;V_2|Q) - I(Z;{V_1}|Q)- I(Z;{V_2}|Q)-\min\{\xi_1,\xi_2\}$ \\
  \rowcolor{gray!10}
  $R_1 + L_2$ & $I(Y_1,X;U_1,U_2|Q) - I(Z;V_2|Q)$ & $I(Y_1;U_1|U_2,Q) +I(X;V_2|Q) - I(Z;V_2|Q)$ \\
  $R_2 + L_1$ & $I(X,Y_2;U_1,U_2|Q) - I(Z;V_1|Q)$ & $I(Y_2;U_2|U_1,Q) +I(X;V_1|Q) - I(Z;V_1|Q)$ \\
  \rowcolor{gray!10}
  $D$ & $\mathbb{E}[d(X,\hat{X}(U_1,U_2,Q))]$ & $\mathbb{E}[d(X,\hat{X}(U_1,U_2,Q))]$ \\
  \hline
\end{tabular}
\vspace{-4mm}
\end{table*}

Next, we provide an outer bound on the rate-distortion-leakage region of the CEO problem under a general distortion measure, which is one of the main contributions in this paper. For $k \in \{1,2\}$, we define the index
\vspace{-2mm}
\begin{align}
    k' = 3-k, \label{kdash}
\end{align}
and an information quantity
\vspace{-2mm}
\begin{align}
    \xi_k = I(V_k;U_{k'}|Y_k,Y_{k'},Q). \label{xi2-conv}
\end{align}
This quantity appears in the constraints of privacy-leakage rate and their sum rate in the outer bound for general distortion measure. The reader may refer to Table \ref{table-g} at the top of the next page for a comparison of each rate constraint between the inner and outer bounds. Note that if the auxiliary random variables $V_1,U_1,U_2,V_2$ satisfy the long Markov chain in \eqref{jdist-inner}, the quantity $\xi_k$ is just zero.
\begin{Theorem} \label{th-conv}
An outer bound $\mathcal{R}_{\rm out}$ {for a general distortion measure} is provided as the closure of the set of all rate tuples $(R_1,R_2,L_1,L_2,D)\in \mathbb{R}^5_+$ s.t.
\begin{align}
    R_1 &\ge I(Y_1;U_1|U_2,Q),~~~R_2 \ge I(Y_2;U_2|U_1,Q), \label{r1-r2-outer} \\
    R_1 + R_2 &\ge I(Y_1,Y_2;U_1,U_2|Q), \label{r1r2-outer} \\
    L_1 &\ge I(X;V_1|V_2,Q) \nonumber \\
    &~~~~+ I(V_1;U_2|Q)-I(Z;V_1|Q) - \xi_1, \nonumber \\
    L_2 &\ge I(X;V_2|V_1,Q) \nonumber \\
    &~~~~+ I(V_2;U_1|Q)-I(Z;V_2|Q) - \xi_2, \nonumber \\
    L_1+L_2 &\ge I(X;V_1,V_2|Q) + I(V_1;V_2|Q) - I(Z;{V_1}|Q) \nonumber \\
    &~~~~ - I(Z;{V_2}|Q)-\min\{\xi_1,\xi_2\}, \nonumber \\
    R_1 + L_2 &\ge I(Y_1;U_1|U_2,Q) +I(X;V_2|Q) - I(Z;V_2|Q), \nonumber \\
    R_2 + L_1 &\ge I(Y_2;U_2|U_1,Q) +I(X;V_1|Q) - I(Z;V_1|Q), \nonumber \\
    D &\ge \mathbb{E}[d(X,\hat{X}(U_1,U_2,Q))], \label{definition2}
\end{align}
where random variables $(V_1,U_1,Y_1,X,Z,Y_2,U_2,V_2)$ satisfy the Markov chain
$
V_k-U_k-Y_k-X-(Y_{k'},Z)
$. The cardinalities of auxiliary random variables can be limited to $|\mathcal{V}_k| \le |\mathcal{Y}_k| + 10$ and $|\mathcal{U}_k| \le (|\mathcal{Y}_k| + 10)(|\mathcal{Y}_k| + 5)$, $|\mathcal{Q}| \le 6$, and a reproduction function $\hat{X}: \mathcal{U}_1\times \mathcal{U}_2\times \mathcal{Q} \longrightarrow \hat{\mathcal{X}}$.
\end{Theorem}
\begin{proof}
The proof is deferred to Appendix \ref{proof-a}. Here, we outline the idea behind our approach. The compression rates and the distortion constraints are derived similarly to those presented in \cite[Appendix C]{naghibis2015}. However, to analyze the privacy-leakage rates and their sum rate, we establish a novel lemma, Lemma \ref{con-jkz}, to deal with their analyses.
\end{proof}

As a special case, when the privacy constraints \eqref{pl} are not considered, i.e., the rate constraints $L_1$, $L_2$, $L_1 + L_2$, $R_1 + L_2$, and $R_2 + L_1$ are removed from Theorem 3, one can check that \eqref{r1-r2-outer}, \eqref{r1r2-outer}, and \eqref{definition2} form the Berger-Tung outer bound \cite{Berger1978}, \cite{tung1978}, \cite[Chapter 12]{GK}.

\section{Statement of Results for Log-Loss Distortion Measure} \label{sect4}
\subsection{Preliminaries}
In this subsection, we define the log-loss distortion measure, which enables the decoder to make soft decisions, with outputs represented as probability distributions \cite{courtade2014,shkel2017}. This distortion measure is an important metric that is widely studied and applied in practical fields such as machine learning and classification tasks. For example, in deep neural networks, it is used in image classification to estimate a distribution that assigns probabilities to different classes \cite{nagnir2018}.
The formal definition of the log-loss distortion measure is given below.

\begin{Definition} \label{log-loss-dis}
Let $\mathcal{P}(\mathcal{X})$ be the set of probability measures in a set $\mathcal{X}$. For $x_t \in \mathcal{X}$ and $\hat{x}_t \in \mathcal{P}(\mathcal{X})$, the symbol-wise distortion in \eqref{dis-equation} is measured by \cite{courtade2014,no-2019}
\begin{align}
d(x_t,\hat{x}_t) = \log\left(\frac{1}{\hat{x}_t(x_t)}\right)~~~~~{\rm for~} t \in [1:n], \label{log-loss-d}
\end{align}
where $\hat{x}_t(\cdot)$ is the $t$-th symbol in the reconstructed sequence $\hat{x}^n$, and this distribution is evaluated based on $x_t$. \qed
\end{Definition}

In Definition \ref{log-loss-dis}, the set $\mathcal{X} = \mathbb{R}$, and $\mathcal{P}(\mathcal{X})$ becomes all possible probability measures in $\mathbb{R}$ for continuous sources \cite{seo2016}.


\subsection{Tight Bound for Discrete Sources Without SI at Eve} \label{tight-bound-ds} 
The model that we deal with in this section corresponds to {Case (II.1) in Section \ref{II-B}, and its joint distribution is given in \eqref{jointd-no-SI}}. For this model, the auxiliary random variable $V_k, k \in \{1,2\},$ does not appear in the expression of the rate-distortion-leakage region, and it is shown to be exactly tight.

For discrete sources, the following inner and outer bounds are obtained. We present the expressions of these bounds in detail so as to facilitate the comprehension of a difference on the rate-distortion-leakage region compared to the model with SI at Eve in the upcoming subsection.
\begin{Proposition} \label{th-log-loss-in}
An inner bound $\mathcal{R}^{\rm L}_{\rm in}$ on the rate-distortion-leakage region of the model without SI at Eve under log-loss distortion is given as the closure of the set of all tuples $(R_1,R_2,L_1,L_2,D) \in \mathbb{R}^5_+$ s.t.
\begin{align}
    R_1 &\ge I(Y_1;U_1|U_2,Q),~~~
    R_2 \ge I(Y_1;U_2|U_1,Q), \nonumber \displaybreak[0] \\
    R_1 + R_2 &\ge I(Y_1,Y_2;U_1,U_2|Q) \nonumber \displaybreak[0] \\
    L_1 &\ge I(X;U_1|U_2,Q),~~~
    L_2 \ge I(X;U_2|U_1,Q), \\
    L_1 + L_2 &\ge I(X;U_1,U_2|Q), \\
    R_1 + L_2 &\ge I(Y_1,X;U_1,U_2|Q), \\
    R_2 + L_1 &\ge I(X,Y_2;U_1,U_2|Q), \\
    D &\ge H(X|U_1,U_2,Q) \label{log-loss-in}
\end{align}
for all joint distributions
\begin{align}
    P_{QXY_1Y_2U_1U_2} = P_{Q}P_{X}P_{Y_1|X}P_{U_1|Y_1Q}P_{Y_2|X}P_{U_2|Y_2Q} \label{2-joint}
\end{align}
with $|\mathcal{U}_k| \le |\mathcal{Y}_k|+3$ and $|Q| \le 6$.
\end{Proposition}
\begin{proof}
The proof can be done by setting $V_k$ and $Z$ as constants and choosing the reproduction function $\hat{X}(U_1,U_2,Q) = P_{X|U_1U_2Q}(x|U_1,U_2,Q)$ for $x \in \mathcal{X}$ in Theorem \ref{th1}.
\end{proof}

To evaluate the distortion constraint in the achievability proof, it suffices to choose $\hat{X}(U_1,U_2,Q) = P_{X|U_1U_2Q}(x|U_1,U_2,Q)$, as this function is a distribution within the set $\mathcal{P}(\mathcal{X})$, satisfying the hypothesis of the definition of log-loss distortion in Definition \ref{log-loss-dis}. Moreover, the resulting bound obtained by this choice aligns with the lower bound derived in the converse proof.

\begin{Proposition} \label{th3}
An outer bound $\mathcal{R}^{\rm L}_{\rm out}$ of the model without SI at Eve is characterized as the closure of the set of all tuples $(R_1,R_2,L_1,L_2,D) \in \mathbb{R}^5_+$ s.t.
\begin{align}
    R_1 &\ge I(Y_1;U_1|X,Q)+H(X|U_2,Q)-D, \label{r1-logloss} \\
    R_2 &\ge I(Y_2;U_2|X,Q)+H(X|U_1,Q)-D, \label{r2-logloss} \\
    R_1 + R_2 &\ge I(Y_1;U_1|X,Q)+I(Y_2;U_2|X,Q) \nonumber \\
    &~~~~+ H(X)-D, \label{r1r2-coverse} \\
    L_1 &\ge H(X|U_2,Q)-D, \\
    L_2 &\ge H(X|U_1,Q)-D, \\
    L_1 + L_2 &\ge H(X)-D, \label{l1l2-prop2} \\
    R_1 + L_2 &\ge I(Y_1;U_1|X,Q)+{H(X)}-D, \\
    R_2 + L_1 &\ge I(Y_2;U_2|X,Q)+{H(X)}-D, \\
    D &\ge H(X|U_1,U_2,Q) \label{theorem3}
\end{align}
for the joint distribution defined in \eqref{2-joint} with $|\mathcal{U}_k| \le |\mathcal{Y}_k|+3$ and $|Q| \le 6$.
\end{Proposition}
\begin{proof}
    See Appendix \ref{apd-B-A}. In the proof, we use a combination of \cite[Lemma 1]{courtade2014} and Lemma \ref{con-lemma1} to analyze each constraint. [27, Lemma 1] is a well-known property for log-loss distortion, which asserts that the expected distortion is lower bounded by a conditional entropy. Lemma 2 is used to simplify the equation developments in the proofs.
\end{proof}

\begin{Theorem} \label{thm-nosi}
The rate-distortion-leakage region of the model without SI at Eve when the distortion is quantified by the log-loss distortion is characterized as
\begin{align}
    \mathcal{R}=\mathcal{R}^{\rm L}_{\rm in}=\mathcal{R}^{\rm L}_{\rm out}.
\end{align}
\end{Theorem}
\begin{proof}
For a detailed discussion, see \cite[Appendix E]{vyo2023}. Apparently, the expression of the inner bound in Proposition \ref{th-log-loss-in} is different from that of the outer bound in Proposition \ref{th3}, but it can be shown that these regions are equivalent by a similar technique used in \cite[Proof of Theorem 3]{courtade2014}. The key point of this technique is to show that $\mathcal{R}^{\rm L}_{\rm in} \supseteq \mathcal{R}^{\rm L}_{\rm out}$ since it is trivial that $\mathcal{R}^{\rm L}_{\rm in} \subseteq \mathcal{R}^{\rm L}_{\rm out}$. To prove that $\mathcal{R}^{\rm L}_{\rm in} \supseteq \mathcal{R}^{\rm L}_{\rm out}$, we first demonstrate that corner points on the dominant face of $\mathcal{R}^{\rm L}_{\rm out}$ for given $(U_1, U_2)$ are always dominated by a rate point in $\mathcal{R}^{\rm L}_{\rm in}$. Then, leveraging the fact that any point on the dominant face can be achieved by time sharing among corner points, we conclude that any point on the dominant face is dominated by a point in $\mathcal{R}^{\rm L}_{\rm in}$.
\end{proof}

In \cite{courtade2014}, the rate-distortion region for the CEO problem without privacy constraints was given under the log-loss distortion measure. It can be verified that, as a special case of Proposition \ref{th3}, the constraints \eqref{r1-logloss}, \eqref{r2-logloss}, \eqref{r1r2-coverse}, and \eqref{theorem3} yield the same expression as in \cite[Theorem 3]{courtade2014}.

\subsection{Tight Bound for Gaussian Sources Without SI at Eve} \label{tight-bound-gs} 
In this subsection, we first derive a tight bound for Gaussian sources and channels, corresponding to Case (II.2) in Section \ref{II-B}, and then provide a simple calculation on the bound. The joint distribution of the model is given in \eqref{jointd-no-SI}, similar to the previous subsection. Consider that each symbol of the sources $X \sim \mathcal{N}(0,\sigma^2_X)$, a Gaussian distribution with zero mean and variance $\sigma^2_X$, and the channels $P_{Y_1|X}$ and $P_{Y_2|X}$ are respectively modeled as $Y_1 = X + N_1$ and $Y_2 = X + N_2$, where the noises $N_1 \sim \mathcal{N}(0,\sigma^2_{N_1})$ and $N_2 \sim \mathcal{N}(0,\sigma^2_{N_2})$.


\begin{Theorem} \label{th4}
The region $\mathcal{R}$ {of the model without SI at Eve} for Gaussian sources under log-loss distortion measure is characterized by the set of all tuples $(R_1,R_2,L_1,L_2,D) \in \mathbb{R}^5_+$ satisfying
\begin{align}
&R_1 \ge r_1 + \frac{1}{2}\log2 \pi e\left(\frac{1}{\sigma^2_X} + \frac{1-2^{-2r_2}}{\sigma^2_{N_2}}\right)^{-1} - D, \label{r1-ll1} \displaybreak[0] \\
&R_2 \ge r_2 + \frac{1}{2}\log2 \pi e\left(\frac{1}{\sigma^2_X} + \frac{1-2^{-2r_1}}{\sigma^2_{N_1}}\right)^{-1} - D, \label{r2-ll1} \displaybreak[0] \\
&R_1 + R_2 \ge r_1 + r_2 + \frac{1}{2}\log (2\pi e \sigma^2_X) - D, \label{r1r2-ll1} \displaybreak[0] \\
&L_1 \ge  \frac{1}{2}\log 2 \pi e\left(\frac{1}{\sigma^2_X} + \frac{1-2^{-2r_2}}{\sigma^2_{N_2}}\right)^{-1} - D, \label{k1-ll1} \displaybreak[0] \\
&L_2 \ge \frac{1}{2}\log2 \pi e \left(\frac{1}{\sigma^2_X} + \frac{1-2^{-2r_1}}{\sigma^2_{N_1}}\right)^{-1} - D, \label{k2-ll1} \\
&L_1 + L_2 \ge \frac{1}{2}\log2 \pi e \sigma^2_{X} - D, \label{k1k2-ll1} \\
&R_1 + L_2 \ge r_1 + \frac{1}{2}\log (2\pi e \sigma^2_X) - D, \label{r1k2-ll1} \\
&R_2 + L_1 \ge r_2 + \frac{1}{2}\log (2\pi e \sigma^2_X) - D \label{wt-si-ll}
\end{align}
for some $r_1,r_2 \ge 0$ {s.t.}
\begin{align}
    D \ge \frac{1}{2}\log 2 \pi e\left( \frac{1}{\sigma^2_X} + \sum_{k=1}^2\frac{1-2^{-2r_k}}{\sigma^2_{N_k}}\right)^{-1}. \label{d-ll1}
\end{align}
\end{Theorem}
\begin{proof}
See Appendix \ref{apd-B-B}. The proof consists of two parts: achievability and converse parts. The achievability part can be proved by considering $U_k = Y_k + \Theta_k$, where $\Theta_k$ represents the noise of the test channel $P_{U_k|Y_k}$. For the converse part, the technique developed in \cite{wit2009,vyo2022-isit} is employed. Specifically, we derive the outer region of Theorem \ref{th4} through Proposition \ref{th3} by fixing $H(X|U_{\mathcal{S}^c},Q), \mathcal{S}^c \subseteq \{1,2\},$ and applying the conditional entropy power inequality \cite[Lemma II]{bergmans1974} to bound each constraint in Proposition \ref{th3}.
\end{proof}

In case the privacy-leakage rates are not considered, i.e., the constraints related to $L_1$ and $L_2$ can be removed from \eqref{wt-si-ll}, Theorem \ref{th4} reduces to the rate-distortion region of the CEO problem under log-loss distortion derived in \cite[Theorem 5]{seo2016}.

\begin{figure}[!t]
\centering
   \includegraphics[scale=0.5]{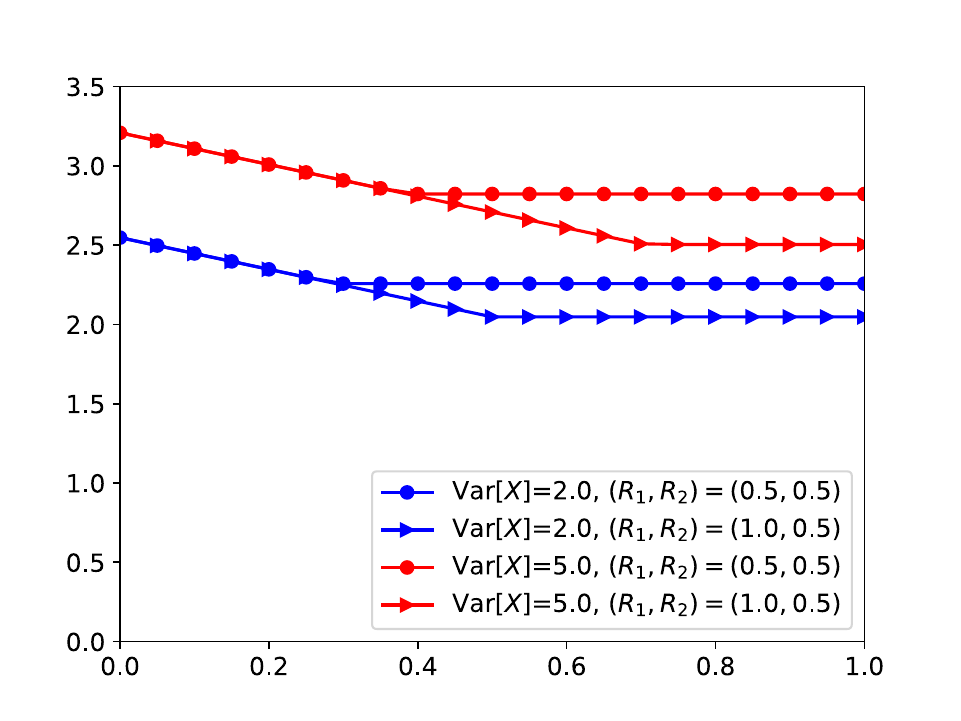}
   \caption{A relation of privacy-leakage rate versus minimum distortion of the Gaussian CEO problem under log-loss distortion measures for the model without SI at Eve. In the calculation for each graph, we set $\sigma^2_{N_1}=\sigma^2_{N_2} = 1$.}
   \label{gscs113}
\end{figure}

Lastly, we conclude this subsection with numerical calculations for Theorem \ref{th4}, examining the relationship between the privacy-leakage rate $L_1$ and the minimum distortion $D$ (min $D$) in a practical scenario where both compression rates $R_1$ and $R_2$ are fixed. Additionally, the constraint of privacy-leakage rate $L_2$ is relaxed in the calculations, that is, \eqref{k2-ll1}, \eqref{k1k2-ll1}, and \eqref{r1k2-ll1} are removed from Theorem \ref{th4}. The detailed numerical results are shown in Figure \ref{gscs113}.

From the behaviors of each graph, one can see that when the privacy-leakage rate $L_1$ becomes large, the minimum distortion is in a saturation state. This is because when $L_1$ is large, the conditions in \eqref{k1-ll1} and \eqref{wt-si-ll} become redundant, and the minimum value of distortion $D$ subject to other constraints \eqref{r1-ll1}, \eqref{r2-ll1}, \eqref{r1r2-ll1}, and \eqref{d-ll1} remains unchanged, and therefore the saturation occurs. Moreover, from the behaviors of graphs in Figure \ref{gscs113}, it is clear that {for} the rate pair $(R_1,R_2)=(1.0,0.5)$, a lower value in regard to the minimum $D$ is achievable compared to the case of $(R_1,R_2) = (0.5,0.5)$, which implies that the smaller the distortion, the larger the compression and privacy-leakage rates.

\subsection{Inner and Outer Bounds for the Model With SI at Eve} \label{si-at-eve}
In the previous subsection, we have seen that for the model without SI at Eve, the exact rate region under log-loss distortion is derivable for both discrete and Gaussian sources. Unfortunately, a tight bound for the model in which SI is available in Eve has not yet been established. However, in this section, we demonstrate that the inner and outer bounds for the model with SI at Eve are quite close, differing only by a minor term. This scenario corresponds to Case (II.3) in Section \ref{II-B}, and the joint distribution of the model is given in \eqref{jointd}. The inner bound is given below.

\begin{Proposition} \label{th-in-ll}
An inner bound $\mathcal{R}^{\rm L}_{\rm in}$ on the rate-distortion-leakage region of the model with SI at Eve for log-loss distortion is given by all constraints in $\mathcal{R}_{\rm in}$ defined in Theorem \ref{th1} with replacing the distortion constraint \eqref{dis-cons} by
\begin{align}
    D &\ge H(X|U_1,U_2,Q),
\end{align}
where auxiliary random variables $V_k,U_k$ for $k \in \{1,2\}$ satisfy the {joint} distribution \eqref{jdist-inner}
with their cardinalities bounded as $|\mathcal{V}_k| \le |\mathcal{Y}_k| + 8$, $|\mathcal{U}_k| \le (|\mathcal{Y}_k| + 8)(|\mathcal{Y}_k| + 6)$, and $|\mathcal{Q}| \le 6$.
\end{Proposition}
\begin{proof} The proof follows from Theorem \ref{th1} by considering $\hat{X}(U_1,U_2,Q) = P_{X|U_1U_2Q}(x|U_1,U_2,Q)$ for $x \in \mathcal{X}$.
\end{proof}

\begin{table*}[t]
\centering
\caption{Comparison of Rate Constraints in Inner and Outer Bounds For Log-loss Distortion Measure}
\label{table-l}
\vspace{-2mm}
\begin{tabular}{| m{1cm} | m{7cm}| m{7cm} | } 
  \hline
    \hfil Rates& \hfil Inner bound & \hfil Outer bound \\ 
   \hline
  \rowcolor{gray!10}
   $R_1$ & $I(Y_1;U_1|U_2,Q)$ & $I(Y_1;U_1|X,Q)+H(X|U_2,Q)-D$ \\ 
  $R_2$ & $I(Y_2;U_2|U_1,Q)$ & $I(Y_2;U_2|X,Q)+H(X|U_1,Q)-D$ \\ 
  \rowcolor{gray!10}
  $R_1 + R_2$ & $I(Y_1,Y_2;U_1,U_2|Q)$ & $I(Y_1;U_1|X,Q)+I(Y_2;U_2|X,Q) + H(X)-D$ \\ 
  $L_1$ & $I(X;U_1|U_2,Q) + I(V_1;U_2|Q)-I(Z;V_1|Q)$ & $H(X|U_2,Q)-D + I(V_1;U_2|Q)-I(Z;V_1|Q)-\xi'$ \\
  \rowcolor{gray!10}
  $L_2$ & $I(X;U_2|U_1,Q) + I(V_2;U_1|Q)-I(Z;V_2|Q)$ & $H(X|U_1,Q)-D + I(V_2;U_1|Q)-I(Z;V_2|Q) -\xi'$ \\
  $L_1 + L_2$ & $I(X;U_1,U_2|Q) + I(V_1;V_2|Q)-I(Z;V_1|Q) -I(Z;V_2|Q)$ & $H(X)-D + I(V_1;V_2|Q)-I(Z;V_1|Q) - I(Z;V_2|Q)-\xi'$ \\
  \rowcolor{gray!10}
  $R_1 + L_2$ & $I(Y_1,X;U_1,U_2|Q) - I(Z;V_2|Q)$ & $I(Y_1;U_1|X,Q)+H(X)-D - I(Z;V_2|Q)$ \\
  $R_2 + L_1$ & $I(X,Y_2;U_1,U_2|Q) - I(Z;V_1|Q)$ & $I(Y_2;U_2|X,Q)+H(X)-D - I(Z;V_1|Q)$ \\
  \rowcolor{gray!10}
  $D$ & $H(X|U_1,U_2,Q)$ & $H(X|U_1,U_2,Q)$ \\
  \hline
\end{tabular}
\vspace{-4mm}
\end{table*}

\begin{Remark}
Upon comparing Propositions \ref{th-log-loss-in} and \ref{th-in-ll}, one can observe that the difference between these regions appears in the constraints related to the privacy-leakage rates. For example, the difference between the privacy-leakage rate $L_1$ of the first agent for the model with and without SI at Eve is $I(V_1;U_2|Q)-I(Z;V_1|Q)$. This quantity indeed serves as an upper bound on the term $-I(Z^n;J_1)$ when evaluating the privacy-leakage rate in the achievability part, as it can be expanded as $I(X^n;J_k|Z^n) = I(X^n;J_k)-I(Z^n;J_k)$.
\end{Remark}

Hereafter, we exhibit an outer bound on the rate-distortion-leakage region in the context of log-loss distortion. A genuine challenge for deriving the outer bound lies in providing the optimal lower bound for the term $-I(Z^n;J_k)$, that is, it should be lower bounded by $I(V_1;U_2|Q)-I(Z;V_1|Q)$ {when the blocklength is sufficiently large}. In the proof, while {the optimal lower} bound for this term has not been achieved, a lower bound that is close to the desired upper one is derived.

To present an outer bound of the rate-distortion-leakage region under log-loss distortion, let us define
$
    \xi' = I(V_1;V_2|Q).
$
This quantity represents the minor term that appears in the constraints of privacy-leakage rate and their sum rate in the outer bound. See Table \ref{table-l} at the top of this page for a comparison of each rate constraint between the inner and outer bounds.

\begin{Proposition} \label{th-out-ll}
An outer bound $\mathcal{R}^{\rm L}_{\rm out}$ of the model with SI at Eve for log-loss distortion is given as the closure of the set of all rate tuples $(R_1,R_2,L_1,L_2,D)\in \mathbb{R}^5_+$ s.t.
\vspace{-2mm}
\begin{align}
    R_1 &\ge I(Y_1;U_1|X,Q)+H(X|U_2,Q)-D,\nonumber \\
    R_2 &\ge I(Y_2;U_2|X,Q)+H(X|U_1,Q)-D, \nonumber \\
    R_1 + R_2 &\ge I(Y_1;U_1|X,Q)+I(Y_2;U_2|X,Q) + {H(X)}-D, \nonumber \\
    L_1 &\ge H(X|U_2,Q)-D + I(V_1;U_2|Q)\nonumber \\
    &~~~~ -I(Z;V_1|Q)-\xi', \nonumber \\
    L_2 &\ge H(X|U_1,Q)-D + I(V_2;U_1|Q)\nonumber \\
    &~~~~ -I(Z;V_2|Q) -\xi', \nonumber \\
    L_1+L_2 &\ge H(X)-D + I(V_1;V_2|Q)-I(Z;V_1|Q) \nonumber \\
    &~~~~ - I(Z;V_2|Q)-\xi', \nonumber \\
    R_1 + L_2 &\ge I(Y_1;U_1|X,Q)+H(X)-D - I(Z;V_2|Q), \nonumber \\
    R_2 + L_1 &\ge I(Y_2;U_2|X,Q)+H(X)-D - I(Z;V_1|Q), \nonumber \\
    D &\ge H(X|U_1,U_2,Q), \label{definition2-ll}
\end{align}
where auxiliary random variables $V_k,U_k$ for $k \in \{1,2\}$ satisfy the {joint} distribution \eqref{jdist-inner}
with $|\mathcal{V}_k| \le |\mathcal{Y}_k| + 9$, $|\mathcal{U}_k| \le (|\mathcal{Y}_k| + 9)(|\mathcal{Y}_k| + 4)$, and $|\mathcal{Q}| \le 6$.
\end{Proposition}
\begin{proof}
See Appendix \ref{proof-out-ll}. The analysis of compression rates and distortion is similar to the arguments used in the derivation of Proposition \ref{th3}. Instead, bounding the privacy constraints requires Lemma \ref{lemma-ll-eve}, which extends Lemma \ref{con-jkz}. The extended lemma plays a key role in deriving lower bounds on the privacy-leakage rates and their sum rate when the distortion is quantified by the log-loss distortion measure.
\end{proof}

The established outer bound in Proposition \ref{th-out-ll} is not yet tight, which can be confirmed when examining the scenario where the distortion is optimal, namely, when $D = H(X|U_1,U_2,Q)$. For this case, it is evident that $\mathcal{R}^{\rm L}_{\rm in} \subseteq \mathcal{R}^{\rm L}_{\rm out}$, but the reverse inclusion does not hold due to the inclusion of $\xi'$ in the constraints of the privacy-leakage rates and their sum rate. Consequently, the characterization of the tight bound for the CEO problem, in which Eve possesses SI under log-loss distortion, remains an open question. However, the inner and outer bounds coincide only in the specific case where the distortion is large, as stated in the following corollary.
\begin{Corollary}
When the distortion $D \ge I(Y_1;U_1|X,Q) + I(Y_2;U_2|X,Q) + H(X)$, the inner and outer bounds coincide.
\end{Corollary}
\begin{proof}
To prove the above corollary, it suffices to show that there is always a rate tuple in the inner bound that dominates any rate tuple with
\begin{align}
D \ge I(Y_1;U_1|X,Q) + I(Y_2;U_2|X,Q) + H(X) \label{coro-D}
\end{align}
in the outer bound. Note that in Proposition \ref{th-out-ll} (outer bound), when the distortion satisfies \eqref{coro-D}, the other rates $R_1,R_2,L_1$, and $L_2$ become zero. Moreover, observe that all rate tuples of the form $(0,0,0,0,D)$ are dominated by the rate tuple $(0,0,0,0,H(X))$ in the inner bound. The latter is achieved by setting all the auxiliary random variables $(V_1,U_1,U_2,V_2)$ in Proposition \ref{th-in-ll} to constants.
\end{proof}

\begin{figure}[!t]
\centering
   \includegraphics[scale=0.6]{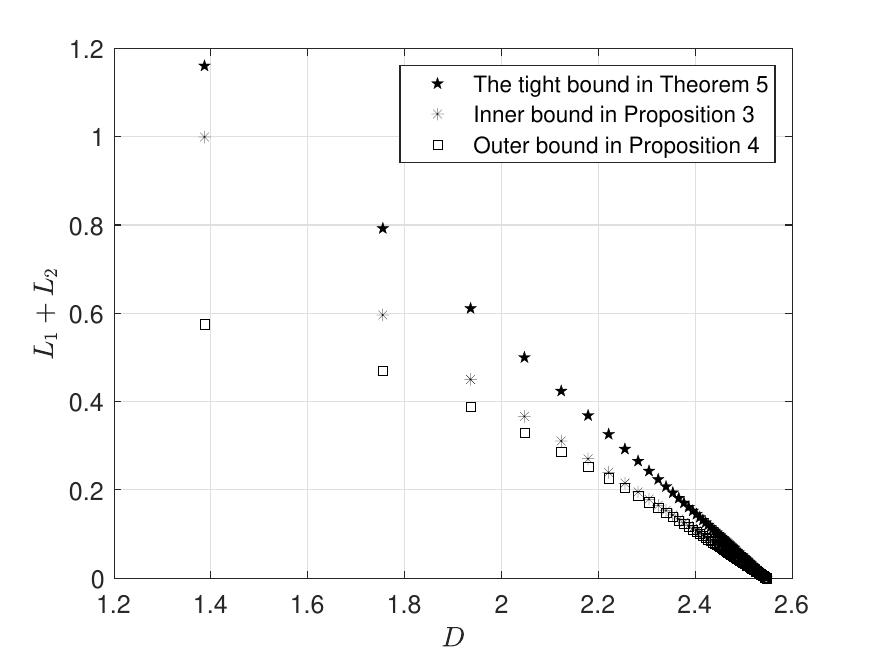}
   \vspace{-5mm}
   \caption{An illustration on the behavior of $L_1 + L_2$ and $D$ of the inner bound (Proposition \ref{th-in-ll}) and the outer bound (Proposition \ref{th-out-ll}) is shown using $*$ and {\tiny $\square$}, respectively. The relation of \eqref{k1k2-ll1} and \eqref{d-ll1} in Theorem \ref{th4} is indicated by $\star$. The right-upper parts of each graph represent the achievable regions.}
   \label{l1l2-D-fig}
   \vspace{-4mm}
\end{figure}

We end this subsection by presenting a numerical example that illustrates a relation between\footnote{{Note that we focus solely on discussing the constraints $L_1 + L_2$ and $D$ to provide a numerical example, and the complete parametric forms of the inner and outer bounds have not yet been fully derived.}} $L_1 + L_2$ and $D$ of the inner and outer bounds derived in this paper. In this calculation, we consider $X \sim \mathcal{N}(0,2)$, $Y_k = X + N_k$, $Z = X + N_Z$, where $N_k \sim \mathcal{N}(0,1)$ and $N_Z \sim \mathcal{N}(0,2)$. For these parameters, when the auxiliary random variables $V_1,U_1,U_2,V_2$ are independent of $(X,Y_1,Y_2,Z)$ and each other, the minimum value of $D$ is $H(X) = 2.547$. As shown in Fig. \ref{l1l2-D-fig}, the gap between the inner and outer bounds narrows as the distortion increases. This occurs because a higher distortion corresponds to a lower correlation between $X$ and $U_1$, as well as between $X$ and $U_2$, which leads to a lower correlation between $U_1$ and $U_2$. Consequently, the correlation between $V_1$ and $V_2$ also weakens, causing the minor term $\xi' = I(V_1;V_2|Q)$ in the outer bound to approach zero. When $D = 2.547$, where $\xi' = 0$, the inner and outer bounds match.

\section{{Conclusions and Open Problems}} \label{sect5}
In this work, we have investigated an outer bound on the rate-distortion-leakage region for the CEO problem in the presence of Eve. As a special case where the log-loss distortion measure is used at the CEO, we have shown that the inner and outer bounds of the model without SI at Eve are exactly tight. Also, we have derived an outer bound for the model where SI is available at Eve and shown that the outer bound is close to the inner bound but not yet tight in general.

In the following, we state some possible future directions.
\begin{itemize}
\item A straightforward extension is to investigate if it is possible to characterize the tight bound on the rate-distortion-leakage region for the model with SI at Eve, which remains unsolved for both quadratic and log-loss distortion measures. For quadratic distortion, an inner bound on the rate-distortion-equivocation region was derived in \cite{naghibis2015}. By the same analogy as for deriving Theorem \ref{th1}, one can easily obtain an inner bound on the rate-distortion-leakage region as well. Thus, the main task is to derive an outer bound on the region that matches the inner bound. For the log-loss distortion, one may need to select different choices of auxiliary random variables that achieve a better lower bound than the one derived in Lemma \ref{lemma-ll-eve}.
\item For the model without SI at Eve, it was demonstrated that the tight bound is derivable for the Gaussian CEO problem when distortion is quantified by the log-loss distortion measure, as shown in Section \ref{tight-bound-gs}. When privacy constraints are not considered, the tight bound for the binary CEO problem was also characterized using the same distortion measure \cite{nagnir2018}. It would be of interest to investigate whether the inner and outer bounds coincide for the binary CEO problem with privacy constraints. Additionally, this discussion could be extended to scenarios where distortion is measured using the Hamming-distance distortion.
\item The model considered in this work focuses on the two-link CEO problem (two agents). As the network becomes more distributed, some practical applications, such as sensor networks, may involve more than two agents \cite{Chen2004}. Therefore, it is a natural extension to study the rate-distortion-leakage region of the CEO problem with $m~(m \ge 2)$ links.
\item Another intriguing research topic is to study the rate-distortion-leakage region in a model where both Eve and the decoder have side information, as explored in \cite{tu2019-tit,ugur2020}. Having side information allows the decoder to gain an advantage over Eve, which could help in characterizing the tight bound under certain conditions, such as when the channel to the decoder is less noisy than that of Eve.
\end{itemize}

\section*{Acknowledgment}
The authors are indebted to Prof. Yasutada Oohama at the University of Electro-Communications for constructive discussions {as well as} providing the references \cite{Berger1978} and \cite{tung1978}. They also thank Prof. Hirosuke Yamamoto at the University of Tokyo for {pointing out} the references \cite{yamamoto1980} and \cite{FlynnGray1978}.



\appendices
\section{Proof of Theorem \ref{th-conv}} \label{proof-a}
Assume that a rate tuple $(R_1,R_2,L_1,L_2,D) \in \mathbb{R}^5_+$ is achievable with respect to Definition \ref{defsystem}, which implies that there exist encoders and decoders that satisfy all the conditions in Definition \ref{defsystem}. As in \cite{naghibis2015}, define $V_{k,t} = (J_k,X^{t-1})$ and $U_{k,t} = (J_k,Y^{t-1}_k,Y^{t-1}_{k'},X^{t-1})$ for $k \in \{1,2\}$ and $t \in [1:n]$, where the definition of $k'$ follows \eqref{kdash}. These settings guarantee the Markov chain
\begin{align}
V_{k,t}-U_{k,t}-Y_{k,t}-{X_t}-(Y_{k',t},Z_t). \label{con-markov}
\end{align}

Prior to the detailed proof, we introduce a lemma used to derive the single-letter expression for the outer bound on the rate-distortion-leakage region, which serves as a key contribution of this proof.
\begin{Lemma} \label{con-jkz} Define $\xi_{k,t} = I(V_{k,t};U_{k',t}|Y_{k,t},Y_{k',t})$. Then, we have that
\begin{align}
    I(J_k;J_{k'}) - I(Z^n;J_k) &\ge \sum_{t=1}^n\{I(V_{k,t};U_{k',t}) -I(Z_t;V_{k,t}) \nonumber \\
    &~~~~~~~~ - \xi_{k,t}\}.
\end{align}
\end{Lemma}
\begin{proof} Observe that
\begin{align}
    &I(J_k;J_{k'}) - I(Z^n;J_k) = H(J_k|Z^n) - H(J_k|J_{k'}) \nonumber \\
    &= H(Y^n_k,J_k|Z^n) - H(Y^n_k|J_k,Z^n) - H(Y^n_k,Y^n_{k'},J_k|J_{k'}) \nonumber \\
    &~~~~+ H(Y^n_k,Y^n_{k'}|J_k,J_{k'}) \nonumber \\
    &= H(Y^n_k|Z^n) - H(Y^n_k|J_k,Z^n) - H(Y^n_k,Y^n_{k'}|J_{k'}) \nonumber \\
    &~~~~ + H(Y^n_k,Y^n_{k'}|J_k,J_{k'}) \label{ttt-va}\\
    &\overset{\rm (a)}= \sum_{t=1}^n\{H(Y_{k,t}|Z_t) - H(Y_{k,t}|J_k,X^{t-1},Y^{t-1}_k,Z^n) \nonumber \\
    &~~~~ - H(Y_{k,t},Y_{k',t}|J_{k'},X^{t-1},Y^{t-1}_k,Y^{t-1}_{k'})\nonumber \\
    &~~~~ + H(Y_{k,t},Y_{k',t}|J_k,J_{k'},X^{t-1},Y^{t-1}_k,Y^{t-1}_{k'})\} \nonumber \\
    &\overset{\rm (b)}\ge \sum_{t=1}^n\{H(Y_{k,t}|Z_t) - H(Y_{k,t}|J_k,X^{t-1},Z_t) \nonumber \\
    &~~~~ - H(Y_{k,t},Y_{k',t}|U_{k',t}) + H(Y_{k,t},Y_{k',t}|V_{k,t},U_{k',t})\} \label{lemma1-b} \displaybreak[0] \\
    &= \sum_{t=1}^n\{H(Y_{k,t}|Z_t) - H(Y_{k,t}|V_{k,t},Z_t) \nonumber \\
    &~~~~- H(Y_{k,t},Y_{k',t}|U_{k',t}) + H(Y_{k,t},Y_{k',t}|V_{k,t},U_{k',t})\} \nonumber \displaybreak[0] \\
    &= \sum_{t=1}^n\{I(Y_{k,t};V_{k,t}|Z_t) - H(Y_{k,t}|U_{k',t}) \nonumber \\
    &~~~~ + H(Y_{k,t}|V_{k,t},U_{k',t})- H(Y_{k',t}|Y_{k,t},U_{k',t}) \nonumber \\
    &~~~~ + H(Y_{k',t}|Y_{k,t},V_{k,t},U_{k',t})\} \nonumber \displaybreak[0] \\
    &= \sum_{t=1}^n\{I(Y_{k,t};V_{k,t}|Z_t) - I(V_{k,t};Y_{k,t}|U_{k',t}) \nonumber \\
    &~~~~ -I(V_{k,t};Y_{k',t}|Y_{k,t},U_{k',t})\} \nonumber \displaybreak[0] \\
    &\overset{\rm (c)}= \sum_{t=1}^n\{H(V_{k,t}|Z_t) - H(V_{k,t}|Y_{k,t}) - H(V_{k,t}|U_{k',t}) \nonumber \\
    &~~~~ + H(V_{k,t}|Y_{k,t},U_{k',t})  -I(V_{k,t};Y_{k',t}|Y_{k,t},U_{k',t})\} \nonumber \displaybreak[0] \\
    &= \sum_{t=1}^n\{H(V_{k,t}|Z_t) - H(V_{k,t}|U_{k',t}) - I(V_{k,t};U_{k',t}|Y_{k,t}) \nonumber \displaybreak[0] \\
    &~~~~ -I(V_{k,t};Y_{k',t}|Y_{k,t},U_{k',t})\} \nonumber \\
    &= \sum_{t=1}^n\{I(V_{k,t};U_{k',t})-I(Z_t;V_{k,t}) \nonumber \\
    &~~~~ - I(V_{k,t};Y_{k',t},U_{k',t}|Y_{k,t})\} \nonumber \\
    &\overset{\rm (d)}= \sum_{t=1}^n\{I(V_{k,t};U_{k',t})-I(Z_t;V_{k,t}) - \xi_{k,t}\}, \label{xi-k}
\end{align}
where (a) is due to the Markov chains $Y_{k,t}-(J_k,Y^{t-1}_k,Z^n)-X^{t-1}$ and $(Y_{k,t},Y_{k',t})-(J_k,J_{k'},Y^{t-1}_k,Y^{t-1}_{k'})-X^{t-1}$, (b) follows as conditioning reduces entropy, and (c) and (d) hold by the Markov chains $V_{k,t}-Y_{k,t}-Z_t$ and $V_{k,t}-Y_{k,t}-Y_{k',t}$, respectively.
\end{proof}

For compression rates and the distortion constraint, the discussions are omitted as they follow the steps demonstrated in \cite[Appendix C]{naghibis2015}. That is, for any $\mathcal{S} \subseteq \{1,2\}$, it holds that
\vspace{-2mm}
\begin{align}
    n(\sum_{k \in \mathcal{S}}R_k + |\mathcal{S}|\delta) &\ge H(J_{\mathcal{S}}) \ge \sum_{t=1}^nI(Y_{\mathcal{S},t};U_{\mathcal{S},t}|U_{\mathcal{S}^c,t}), \label{r2-con} \\
    n(D + \delta) &\ge \sum_{t=1}^n\mathbb{E}[d(X_t,\hat{X}_t(U_{k,t},U_{k',t}))].
    \label{result-2015}
\end{align}

The analyses of the privacy-leakage rates, their sum rate, and the mixed sum rate are verified as follows: 

\medskip
\noindent{\em {\em  i)} Analysis of the Privacy-Leakage Rate}: From \eqref{pl}, we have
\vspace{-2mm}
\begin{align}
    &n(L_k + \delta) \ge I(X^n;J_k|Z^n) \overset{\rm (a)}= I(X^n;J_k) - I(Z^n;J_k) \label{lk-delta} \\
    &\overset{\rm (b)}= I(X^n;J_k|J_{k'}) + I(J_{k'};J_k) - I(Z^n;J_k) \nonumber \displaybreak[0] \\
    &= \sum_{t=1}^nI(X_t;J_k|J_{k'},X^{t-1}) + I(J_{k'};J_k) - I(Z^n;J_k) \nonumber \displaybreak[0] \\
    &=\sum_{t=1}^nI(X_t;J_k,X^{t-1}|J_{k'},X^{t-1}) + I(J_{k'};J_k) - I(Z^n;J_k) \nonumber \displaybreak[0] \\
    &\overset{\rm (c)} \ge \sum_{t=1}^n\{I(X_t;V_{k,t}|V_{k',t})+I(V_{k,t};U_{k',t})-I(Z_t;V_{k,t}) \nonumber \\
    &~~~~ - \xi_{k,t}\},
\end{align}
where (a) and (b) are due to the Markov chains $J_k-X^n-Z^n$ and $J_k-X^n-J_{k'}$, respectively, and (c) follows {since} Lemma \ref{con-jkz} applies.

\medskip
\noindent{\em {\em  ii)} Analysis of the Sum Rate}: Again from \eqref{pl}, on one hand, observe that
\vspace{-2mm}
\begin{align}
    &n(L_1 + L_2 + 2\delta) \ge I(X^n;J_1|Z^n) + I(X^n;J_2|Z^n) \nonumber \\
    &=I(X^n;J_1) + I(X^n;J_2) -I(Z^n;J_1) - I(Z^n;J_2) \label{tau-1} \displaybreak[0] \\
    &=I(X^n;J_1) + I(X^n;J_2|J_1) + I(J_1;J_2)-I(Z^n;J_1) \nonumber \\
    &~~~~ - I(Z^n;J_2) \nonumber  \displaybreak[0] \\
    &\overset{\rm (a)}\ge \sum_{t=1}^n\{I(X_t;J_1,J_2,X^{t-1})- I(Z_t;J_2,X^{t-1},Z^{t-1})\} \nonumber \\
    &~~~~ + I(J_1;J_2)-I(Z^n;J_1) \nonumber \displaybreak[0] \\
    &\overset{\rm (b)}=\sum_{t=1}^n\{I(X_t;V_{1,t},V_{2,t})- I(Z_t;V_{2,t})\} \nonumber \\
    &~~~~+ I(J_1;J_2)-I(Z^n;J_1) \nonumber \displaybreak[0] \\
    &\overset{\rm (c)}\ge \sum_{t=1}^n\{I(X_t;V_{1,t},V_{2,t}) + I(V_{1,t};V_{2,t}) - I(Z_t;V_{1,t}) \nonumber \\
    &~~~~ -I(Z_t;V_{2,t}) - \xi_{1,t}\},
    \label{l1l2-1}
\end{align}
where (a) holds {since} each symbol of $(X^n,Z^n)$ is i.i.d. generated and thus $\sum_{t=1}^nI(X_t;X^{t-1})=\sum_{t=1}^nI(Z_t;Z^{t-1})=0$, (b) is due to the Markov chain $Z_t-(J_2,X^{t-1})-Z^{t-1}$, and (c) is due to Lemma \ref{con-jkz}.

On the other hand, in \eqref{tau-1}, we can also rearrange the equation as
$I(X^n;J_2) + I(X^n;J_1|J_2) -I(Z^n;J_1) + I(J_2;J_1) - I(Z^n;J_2)$. Tracing the steps from \eqref{tau-1} to \eqref{l1l2-1}, we also obtain
\begin{align}
    n&(L_1 + L_2 + 2\delta) \ge \sum_{t=1}^n\{I(X_t;V_{1,t},V_{2,t}) + I(V_{1,t};V_{2,t}) \nonumber \\
    &~~~~ - I(Z_t;V_{1,t})-I(Z_t;V_{2,t}) - \xi_{2,t}\}.
\end{align}
As a result, it follows that
\begin{align}
    n&(L_1 + L_2 + 2\delta) \ge \sum_{t=1}^n\{I(X_t;V_{1,t},V_{2,t}) + I(V_{1,t};V_{2,t}) \nonumber \\
    &~~~~ - I(Z_t;V_{1,t})-I(Z_t;V_{2,t}) - \min\{\xi_{1,t},\xi_{2,t}\}\}. \label{l1l2-2}
\end{align}

\medskip
\noindent{\em {\em iii)} Analysis of the Mixed Sum Rate}: From \eqref{storage} and \eqref{pl},
\begin{align}
    n&(R_{k'} + L_k + 2\delta) \ge \log |\mathcal{J}_{k'}| + I(X^n;J_k|Z^n) \nonumber \\
    &\ge H(J_{k'}) + I(X^n;J_k|Z^n) \label{rkdlk} \\
    &= H(J_{k'}) + I(X^n;J_k) - I(Z^n;J_k) \nonumber \\
    &\overset{\rm (a)}= H(J_{k'}) + \sum_{t=1}^n\{I(X_t;J_k,X^{t-1}) - I(Z_t;J_k,Z^{t-1})\}\nonumber \\
    &\ge H(J_{k'}) + \sum_{t=1}^n\{I(X_t;J_k,X^{t-1}) \nonumber \\
    &~~~~ - I(Z_t;J_k,X^{t-1},Z^{t-1})\}\nonumber \\
    &\overset{\rm (b)}= H(J_{k'}) + \sum_{t=1}^n\{I(X_t;V_t) - I(Z_t;V_t)\}\nonumber \\
    &\overset{\rm (c)}\ge \sum_{t=1}^n\{I(Y_{k',t};U_{k',t}|U_{k,t}) + I(X_t;V_{k,t}) - I(Z_t;V_{k,t})\},
\end{align}
where (a) follows from the i.i.d. property of $(X^n,Z^n)$, (b) holds because the Markov chain $Z_t-(J_k,X^{t-1})-Z^{t-1}$ is applied, and (c) is due to \eqref{r2-con}.

The proof is wrapped up with the standard single-letterization argument by considering a time-sharing random variable $Q \sim~{\rm Unif}[1:n]$ and denoting $X = X_Q$, $Z = Z_Q$, $Y_k = Y_{k,Q}$, $Y_{k'} = Y_{k',Q}$, $U_k = (U_{k,Q},Q)$, and $V_k = (V_{k,Q},Q)$, so that the Markov chain $V_k-U_k-Y_k-X-(Y_{k'},Z)$ holds.

Moreover, using the support lemma \cite[Appendix C]{GK}, one can show that it suffices to limit $|\mathcal{V}_k| \le |\mathcal{Y}_k| + 10$ and $|\mathcal{U}_k| \le (|\mathcal{Y}_k| + 10)(|\mathcal{Y}_k| + 5)$ to preserve $\mathcal{R}_{\rm out}$. Furthermore, $|\mathcal{Q}| \le 6$ follows from the Fenchel-Eggleston-Carath\'eodory theorem \cite[Appendix A]{GK}. Finally, letting $n \rightarrow \infty$ and $\delta \downarrow 0$, the converse proof completes.
\qed
\section{Proof of Proposition \ref{th3} and Theorem \ref{th4}} \label{appendixB}

\subsection{Proof of Proposition \ref{th3}} \label{apd-B-A}

For the converse part, again assume that a rate tuple $(R_1,R_2,L_1,L_2,D) \in \mathbb{R}^5_+$ is achievable with respect to Definition \ref{defsystem}.

For $k \in \{1,2\}$ and $t \in [1:n]$, as in \cite{courtade2014}, we define
$U_{k,t} = (J_k,Y^{t-1}_k)$ and $Q_t = X^{n\backslash t}$. Note that conditioned on $Q_t$, the random variables $(V_{k,t},U_{k,t},Y_{k,t},X_{t},Z_{t})$ satisfy the long Markov chain defined by \eqref{2-joint}.

The next lemma is useful for simplifying equation developments during the analysis. For a set $\mathcal{K} \subseteq \{1,2\}$, the notation $J_{\mathcal{K}} = 0$ for $\mathcal{K}=\emptyset$, $J_{\mathcal{K}} = J_1$ or $J_2$ for $\mathcal{K} = \{1\}$ or $\{2\}$, respectively, and $J_{\mathcal{K}} = (J_1,J_2)$ for $\mathcal{K} = \{1,2\}$.
\begin{Lemma} \label{con-lemma1} It holds that for all $\mathcal{K} \subseteq \{1,2\}$,
\begin{align}
    H(J_\mathcal{K}|X^n) = \sum_{k \in \mathcal{K}}\Big\{\sum_{t=1}^nI(Y_{k,t};U_{k,t}|X_t,Q_t)\Big\}. \label{lemma-yjxn}
\end{align}
\end{Lemma}
\begin{proof} The equation can be developed as
\begin{align}
    H(J_{\mathcal{K}}&|X^n) = H(Y^n_{\mathcal{K}},J_{\mathcal{K}}|X^n) - H(Y^n_{\mathcal{K}}|J_{\mathcal{K}},X^n) \nonumber \\
    &\overset{\rm (a)}= {H(Y^n_{\mathcal{K}}|X^{n}) - H(Y^n_{\mathcal{K}}|J_{\mathcal{K}},X^n)} \nonumber \\
    &= \sum_{t=1}^n\Big\{H(Y_{\mathcal{K},t}|X^{n}) - H(Y_{\mathcal{K},t}|J_\mathcal{K},Y^{t-1}_{\mathcal{K}},X^{n})\Big\} \nonumber \\
    &= \sum_{t=1}^n\Big\{H(Y_{\mathcal{K},t}|X_t,Q_t) - H(Y_{\mathcal{K},t}|U_{\mathcal{K},t},X_t,Q_t)\Big\} \nonumber \\
    &\overset{\rm (b)}= \sum_{t=1}^n\Big\{\sum_{k \in \mathcal{K}}I(Y_{k,t};U_{k,t}|X_t,Q_t)\Big\},
\end{align}
where (a) follows because $J_{\mathcal{K}}$ is a function of $Y^n_{\mathcal{K}}$, and (b) is due to the Markov chain $Y_{k,t}-(X_t,U_{k,t})-({Y_{k',t}},U_{{k'},t})$ for {$k \in \{1,2\}$}.
\end{proof}

Now we start analyzing each constraint. For the compression rates and distortion constraint, it was shown in \cite{courtade2014} that for any $\mathcal{S} \subseteq \{1,2\}$,
\begin{align}
n&(\sum_{k \in \mathcal{S}}R_k +2\delta) \ge {H(J_{\mathcal{S}})} \ge \sum_{t=1}^n\Big\{H(X_t|U_{\mathcal{S}^c,t},Q_t)\Big\} \nonumber \\
&+ \sum_{k \in \mathcal{S}}\Big\{\sum_{t=1}^nI(Y_{k,t};{U_{k,t}}|X_t,Q_t)\Big\} - n(D + \delta), \label{hjsc1111} \\
&{n(D+\delta) \ge \sum_{t=1}^nH(X_t|U_{1,t},U_{2,t},Q_t)}. \label{dhxu1u2q}
\end{align}

Note that for a set $\mathcal{K} \subseteq \mathcal{S}$ {with} $\mathcal{K} \cup \mathcal{K}^c = \mathcal{S}$, the summation $\sum_{k \in \mathcal{S}}\{\sum_{t=1}^nI(Y_{k,t};{U_{k,t}}|X_t,Q_t)\}$ on the right-hand side of \eqref{hjsc1111} can be rewritten as a sum of the following terms
\begin{align}
&\sum_{k \in \mathcal{K}}\Big\{\sum_{t=1}^nI(Y_{k,t};{U_{k,t}}|X_t,Q_t)\Big\} \nonumber \\
&~~~~+ \sum_{k' \in \mathcal{K}^c}\Big\{\sum_{t=1}^nI(Y_{k',t};{U_{k',t}}|X_t,Q_t)\Big\}. \label{hjsc1112}
\end{align}

In the rest of the proof, we verify the constraints of privacy-leakage rates, their sum rate, and the mixed sum rate between privacy-leakage and compression rates. From \eqref{storage} and \eqref{pl} with $Z =\emptyset $, for all $\mathcal{K} \subseteq \mathcal{S} \subseteq \{1,2\}$, we have that
\begin{align}
    &n(\sum_{k \in \mathcal{K}}R_k + \sum_{k' \in \mathcal{K}^c}L_{k'} + 2\delta) \ge \log |\mathcal{J}_{\mathcal{K}}| + \sum_{k' \in \mathcal{K}^c} I(X^n;J_{k'}) \nonumber \\
    &\ge H(J_{\mathcal{K}}) + H(J_{\mathcal{K}^c}) - \sum_{k' \in \mathcal{K}^c}H(J_{k'}|X^n) \nonumber \\
    &\ge H(J_{\mathcal{S}}) - \sum_{k' \in \mathcal{K}^c} H(J_{k'}|X^n) \nonumber \\
    &\overset{\rm (a)}\ge \sum_{t=1}^n\Big\{H(X_t|U_{\mathcal{S}^c,t},Q_t) + \sum_{k \in \mathcal{S}}I(Y_{k,t};{U_{k,t}}|X_t,Q_t) \nonumber \\
    &~~~~ - \sum_{k' \in \mathcal{K}^c}I(Y_{k',t};{U_{k',t}}|X_t,Q_t)\Big\}-n(D + \delta) \nonumber \\
    &\overset{\rm (b)}= \sum_{t=1}^n\Big\{\sum_{k \in \mathcal{K}}I(Y_{k,t};U_{k,t}|X_t,Q_t) + H(X_t|U_{\mathcal{S}^c,t},Q_t)\Big\} \nonumber \\
    &~~~~ -n(D + \delta),\label{lk-logloss}
\end{align}
where (a) is due to \eqref{lemma-yjxn} and \eqref{hjsc1111} and (b) holds since \eqref{hjsc1112} applies. In \eqref{lk-logloss}, the constraint on $L_k$ is derived when taking $\mathcal{S}=\{k\}$ and $\mathcal{K}=\emptyset$. The constraint on $L_1+L_2$ follows from specifying $\mathcal{S}=\{1,2\}$ and $\mathcal{K}=\emptyset$. In a similar manner, by setting $\mathcal{S}=\{k,k'\}$ and $\mathcal{K}=\{k\}$, we have the constraint on $R_k + L_{k'}$.

By the standard argument for single letterization with defining $X = X_T$, $Z = Z_T$, $Y_{k} = Y_{k,T}$, $U_k = U_{k,T}$, and $Q = (T,Q_T)$ for $T \sim~{\rm Unif}[1:n]$, and letting $n \rightarrow \infty$ and $\delta \downarrow 0$, any achievable tuple $(R_1,R_2,L_1,L_2,D)$ of the CEO problem with physical identifiers is contained in $\mathcal{R}_{\rm out}$.

Moreover, using the Fenchel-Eggleston-Carath\'eodory theorem \cite[Appendix A]{GK}, one can limit $|\mathcal{Q}| \le 6$, and by the support lemma \cite[Appendix C]{GK}, we can bound the cardinality on $U_k$ as $|\mathcal{U}_k| \le |\mathcal{Y}_k| + 3$. \qed

\subsection{Proof of Theorem \ref{th4}} \label{apd-B-B}
For achievability proof, consider that for $k \in \{1,2\}$,
$
    U_k = Y_k + \Theta_k,
$
where $\Theta_k \in \mathcal{N}(0,\sigma^2_{\Theta_k})$. The rest of the achievability proof follows standard techniques, and thus we omit the details. We refer the reader to \cite[Appendix F]{naghibis2015} for a similar proof.

\medskip
For the converse part, note that the expression of single-letter characterization in \eqref{theorem3} also holds for Gaussian sources. Therefore, we will use the expression to derive the closed-form expression for the Gaussian case.
Similar to the technique {developed} in \cite{wit2009,vyo2022-isit}, for $k \in \{1,2\}$, let us fix
\begin{align}
    h&(X|U_{\mathcal{S}^c},Q)= \frac{1}{2}\log2 \pi e \Big( \frac{1}{\sigma^2_X} + \sum_{k \in \mathcal{S}^c}\frac{1}{\sigma^2_{N_k} + \beta_k}\Big)^{-1} \nonumber \\
    &= \frac{1}{2}\log2 \pi e \Big( \frac{1}{\sigma^2_X} + \sum_{k \in \mathcal{S}^c}\frac{1-2^{-2r_k}}{\sigma^2_{N_k}}\Big)^{-1},~~~\label{hxu1u2-setting}
\end{align}
where $r_k = \frac{1}{2}\log\Big(\frac{\sigma^2_{N_k} + \beta_k}{\beta_k}\Big)$ for $\beta_k > 0$. Note that $r_k \ge 0$ for any $\beta_k > 0$. When the set $\mathcal{S}^c$ is singleton, {\eqref{hxu1u2-setting}} reduces to
\begin{align}
    h(X|U_k,Q) &= \frac{1}{2}\log2 \pi e \Big( \frac{1}{\sigma^2_X} + \frac{1-2^{-2r_k}}{\sigma^2_{N_k}}\Big)^{-1} \nonumber \\
    &= \frac{1}{2}\log2 \pi e \Big( \frac{\sigma^2_X(\sigma^2_{N_k} + \beta_k)}{\sigma^2_X+\sigma^2_{N_k} + \beta_k}\Big)~~~\label{hxu1u2-setting-k}
\end{align}
for $k \in \{1,2\}$. It is not so difficult to verify that there exists $r_k$ such that \eqref{hxu1u2-setting-k} holds. The setting in \eqref{hxu1u2-setting-k} {stems} from the fact that the term inside the parentheses is a continuous and monotonically decreasing function with respect to $r_k$, and when the auxiliary random variable $U_k$ varies, there always exists a $r_k$ satisfying this condition, since one can see that $\frac{1}{2}\log2 \pi e \big(\frac{\sigma^2_X\sigma^2_{N_k}}{\sigma^2_X + \sigma^2_{N_k}}\big) = h(X|Y_k) \le h(X|U_k,Q) \le h(X)= \frac{1}{2}\log2 \pi e \sigma^2_X$ holds. Taking $r_k \rightarrow \infty~(\beta_k \downarrow 0)$ and $r_k \downarrow 0~(\beta_k \rightarrow \infty)$, we achieve the lower and upper bounds of this relation. Using a similar analogy, we can check that \eqref{hxu1u2-setting} is valid for any set of $\mathcal{S}^c$.

To further bound the right-hand side in \eqref{theorem3}, we need to find the optimal lower bound for the term $I(Y_k;U_k|X,Q)$ as $h(X|U_{\mathcal{S}^c},Q)$ is already fixed in \eqref{hxu1u2-setting}. In particular, the term
$I(Y_k;U_k|X,Q) = I(Y_k;U_k|Q)-I(X;U_k|Q) = h(Y_k)-h(Y_k|U_k,Q) - h(X) + h(X|U_k,Q)$. Since $h(Y_k)$ and $h(X)$ are constant, and $h(X|U_k,Q)$ is obtained from \eqref{hxu1u2-setting-k}, the remaining tasks are to derive the optimal upper bound on $h(Y_k|U_k,Q)$. Following the technique used in \cite{wataoha-tifs2010}, the converted channel from $Y_k$ to $X$ is given as
\begin{align}
    X = \frac{\sigma^2_X}{\sigma^2_X + \sigma^2_{N_k}}Y_k + N'_k,~N'_k \sim \mathcal{N}\Big(0,\frac{\sigma^2_X\sigma^2_{N_k}}{\sigma^2_X + \sigma^2_{N_k}}\Big). \label{xyndash}
\end{align}
Applying the conditional entropy power inequality \cite{bergmans1974} to \eqref{xyndash} {yields}
\begin{align*}
    &2^{2h(X|U_k,Q)}
    \ge 2^{2h(\frac{\sigma^2_X}{\sigma^2_X + \sigma^2_{N_k}}Y_k|U_k,Q)} + 2^{2h(N'_k|U_k,Q)} \nonumber \\
    &= \Big(\frac{\sigma^2_X}{\sigma^2_X + \sigma^2_{N_k}}\Big)^22^{2h(Y_k|U_k,Q)} + 2\pi e \Big(\frac{\sigma^2_X\sigma^2_{N_k}}{\sigma^2_X + \sigma^2_{N_k}}\Big).
\end{align*}
Now using the condition \eqref{hxu1u2-setting-k} and after some algebra, it holds that
\begin{align}
    2^{2h(Y_k|U_k,Q)}
    \le 2 \pi e \Big(\frac{\beta_k(\sigma^2_X + \sigma^2_{N_k})}{\sigma^2_X + \sigma^2_{N_k} + \beta_k}\Big). \label{iykxku}
\end{align}
Therefore,
\begin{align}
    h(Y_k|U_k,Q) \le \frac{1}{2}\log 2 \pi e \Big(\frac{\beta_k(\sigma^2_X + \sigma^2_{N_k})}{\sigma^2_X + \sigma^2_{N_k} + \beta_k}\Big), \label{1-22-33}
\end{align}
and combining the condition in \eqref{hxu1u2-setting-k} with \eqref{1-22-33} yields
\begin{align}
    &I(Y_k;U_k|X,Q) \nonumber \\
    &= h(Y_k)-h(Y_k|U_k,Q) - h(X) + h(X|U_k,Q) \nonumber \\
    &\ge \frac{1}{2}\log 2 \pi e \Big(\sigma^2_X + \sigma^2_{N_k}\Big) - \frac{1}{2}\log 2 \pi e \Big(\frac{\beta_k(\sigma^2_X + \sigma^2_{N_k})}{\sigma^2_X + \sigma^2_{N_k} + \beta_k}\Big) \nonumber \\
    &~~~- \frac{1}{2}\log 2 \pi e\sigma^2_X + \frac{1}{2}\log2 \pi e \Big( \frac{\sigma^2_X(\sigma^2_{N_k} + \beta_k)}{\sigma^2_X+\sigma^2_{N_k} + \beta_k}\Big) \nonumber \\
    &= \frac{1}{2}\log \Big( \frac{\sigma^2_{N_k} + \beta_k}{ \beta_k}\Big) \nonumber \\
    &= r_k. \label{iykukx}
\end{align}

Finally, substituting \eqref{hxu1u2-setting} and \eqref{iykukx} into the constraints in \eqref{r2-logloss}--\eqref{theorem3}, the converse proof of Theorem \ref{th4} is completed.
\qed
\section{Proof of Proposition \ref{th-out-ll}} \label{proof-out-ll}
Assume that a rate tuple $(R_1,R_2,L_1,L_2,D) \in \mathbb{R}^5_+$ is achievable with respect to Definition \ref{defsystem}. Define
\begin{align}
   V_{k,t} = (J_k,Y^{t-1}_k),~U_{k,t} = (J_k,Y^{t-1}_k,X^{t-1}),~Q_t = X^{n\backslash t} \label{vkukqt}
\end{align}
for $k \in \{1,2\}$. Note that conditioned on $Q_t$, the random variables $V_{1,t}$, $U_{1,t}$, $Y_{1,t}$, $X_{t}$, $Z_{t}$, $Y_{2,t}$, $U_{2,t}$, and $V_{2,t}$ satisfy all possible Markov chains defined by \eqref{jdist-inner}.

In the following lemma, we derive a lower bound on the {term} $-I(Z^n;J_k)$ for the settings in \eqref{vkukqt} via Lemma \ref{con-jkz}.

\begin{Lemma} \label{lemma-ll-eve} Define $\xi'_{t} =I(V_{1,t};V_{2,t}|Q_t)$. It holds that
\begin{align}
    -I(Z^n;J_k) \ge\sum_{t=1}^n\{I(V_{k,t};U_{k',t}|Q_t)-I(Z_t;V_{k,t}|Q_t) - \xi'_{t}\}. \label{aa-lemma1}
\end{align}
\end{Lemma}
\begin{proof} From the same development in \eqref{ttt-va}, it follows that
\begin{align}
    I&(J_k;J_{k'}) - I(Z^n;J_k) \nonumber \displaybreak[0] \\
    &= \sum_{t=1}^n\{H(Y_{k,t}|Z_t) - H(Y_{k,t}|J_k,Y^{t-1}_k,Z^n)\nonumber \\
    &~~~~ - H(Y_{k,t},Y_{k',t}|J_{k'},Y^{t-1}_k,Y^{t-1}_{k'})\nonumber \\
    &~~~~+ H(Y_{k,t},Y_{k',t}|J_k,J_{k'},Y^{t-1}_k,Y^{t-1}_{k'})\} \nonumber \displaybreak[0] \\
    &= \sum_{t=1}^n\{H(Y_{k,t}|Z_t) - H(Y_{k,t}|J_k,Y^{t-1}_k,Z^n,Q_t)\nonumber \displaybreak[0]\\
    &~~~~- I(Q_t;Y_{k,t}|J_k,Z^n,Y^{t-1}_k)\nonumber \\
    &~~~~-H(Y_{k,t},Y_{k',t}|J_{k'},Y^{t-1}_k,Y^{t-1}_{k'},Q_t) \nonumber \\
    &~~~~- I(Q_t;Y_{k,t},Y_{k',t}|J_{k'},Y^{t-1}_k,Y^{t-1}_{k'}) \nonumber \\
    &~~~~+ H(Y_{k,t},Y_{k',t}|J_k,J_{k'},Y^{t-1}_k,Y^{t-1}_{k'},Q_t) \nonumber\\
    &~~~~+ I(Q_t;Y_{k,t},Y_{k',t}|J_k,J_{k'},Y^{t-1}_k,Y^{t-1}_{k'})\} \nonumber \displaybreak[0] \\
    &\overset{\rm (a)}\ge \sum_{t=1}^n\{H(Y_{k,t}|Z_t,Q_t) - H(Y_{k,t}|V_{k,t},Z_t,Q_t) \nonumber \\
    &~~~~-H(Y_{k,t},Y_{k',t}|U_{k',t},Q_t) \nonumber \\
    &~~~~+ H(Y_{k,t},Y_{k',t}|V_{k,t},U_{k',t},Q_t) \nonumber \\
    &~~~~- I(Q_t;Y_{k,t}|J_k,Z^n,Y^{t-1}_k) \nonumber \\
    &~~~~- I(Q_t;Y_{k,t},Y_{k',t}|J_{k'},Y^{t-1}_k,Y^{t-1}_{k'}) \nonumber \\
    &~~~~+ I(Q_t;Y_{k,t},Y_{k',t}|J_k,J_{k'},Y^{t-1}_k,Y^{t-1}_{k'})\} \nonumber \\
    &\overset{\rm (b)}\ge \sum_{t=1}^n\{I(V_{k,t};U_{k't}|Q_t)-I(Z_t;V_{k,t}|Q_t){\}} + F \label{aa-a}
\end{align}
with
\begin{align*}
    F &\triangleq \sum_{t=1}^n\{-I(Q_t;Y_{k,t}|J_k,Z^n,Y^{t-1}_k) \nonumber \\
    &~~~~- I(Q_t;Y_{k,t},Y_{k',t}|J_{k'},Y^{t-1}_k,Y^{t-1}_{k'}) \nonumber \\
    &~~~~+ I(Q_t;Y_{k,t},Y_{k',t}|J_k,J_{k'},Y^{t-1}_k,Y^{t-1}_{k'})\},
\end{align*}
where (a) follows since the Markov chain $Y_{k,t}-(J_k,Y^{t-1}_k,Z_t,Q_t)-Z^{n\backslash t}$ is applied and conditioning reduces entropy, and (b) follows by the steps from \eqref{lemma1-b} to \eqref{xi-k} as the conditional mutual information $I(V_{k,t};U_{k',t}|Y_{k,t},Y_{k',t},Q_t) = 0$.

Observe that $F$ can be expressed as in \eqref{lemma-ll} at the top of the next page,
\begin{figure*}[t]
\begin{align}
    F&\overset{\rm (a)}=\sum_{t=1}^n\{- I(X^n_{t+1};Y_{k,t}|J_k,Z^n,Y^{t-1}_k)- I(X^n_{t+1};Y_{k,t},Y_{k',t}|J_{k'},Y^{t-1}_k,Y^{t-1}_{k'})+ I(X^n_{t+1};Y_{k,t},Y_{k',t}|J_k,J_{k'},Y^{t-1}_k,Y^{t-1}_{k'})\} \nonumber \displaybreak[0]\\
    &\overset{\rm (b)}=\sum_{t=1}^n\{-I(X_{t};Y^{t-1}_k|J_k,X^n_{t+1},Z^n) - I(X_{t};Y^{t-1}_k,Y^{t-1}_{k'}|J_{k'},X^n_{t+1}) + I(X_t;Y^{t-1}_k,Y^{t-1}_{k'}|J_k,J_{k'},X^n_{t+1})\} \nonumber \displaybreak[0]\\
    &=\sum_{t=1}^n\{-H(X_{t}|J_k,X^n_{t+1},Z^n) + H(X_{t}|J_k,Y^{t-1}_k,X^n_{t+1},Z^n) - H(X_{t}|J_{k'},X^n_{t+1}) + H(X_{t}|J_{k'},Y^{t-1}_k,Y^{t-1}_{k'},X^n_{t+1})\nonumber \\
    &~~~~+H(X_t|J_k,J_{k'},X^n_{t+1}) - H (X_t|Y^{t-1}_k,Y^{t-1}_{k'},J_k,J_{k'},X^n_{t+1})\} \nonumber \displaybreak[0]\\
    &=\sum_{t=1}^n\{I(X_{t};J_k|X^n_{t+1},Z^n)-H(X_{t}|Z_t) + H(X_{t}|J_k,Y^{t-1}_k,X^n_{t+1},Z^n)+I(X_{t};J_{k'}|X^n_{t+1}) -H(X_t) \nonumber \\
    &~~~~+ H(X_{t}|J_{k'},Y^{t-1}_k,Y^{t-1}_{k'},X^n_{t+1})+I(X_t;Y^{t-1}_{k'}|J_k,J_{k'},X^n_{t+1}) - I(X_t;J_k|J_{k'},Y^{t-1}_{k'},X^n_{t+1}) + H(X_t|J_{k'},Y^{t-1}_{k'},X^n_{t+1}) \nonumber \\
    &~~~~- H(X_t|Y^{t-1}_k,Y^{t-1}_{k'},J_k,J_{k'},X^n_{t+1})\} \nonumber \displaybreak[0]\\
    &\overset{\rm (c)}=\sum_{t=1}^n\{-H(X_{t}|Z_t) + H(X_{t}|J_k,Y^{t-1}_k,X^n_{t+1},Z^n) -H(X_t) + H(X_{t}|J_{k'},Y^{t-1}_k,Y^{t-1}_{k'},X^n_{t+1}) \nonumber \\
    &~~~~ + H(X_t|J_{k'},Y^{t-1},X^n_{t+1}) - H(X_t|Y^{t-1}_k,Y^{t-1}_{k'},J_k,J_{k'},X^n_{t+1})\} + \sum_{t=1}^n\{I(X_{t};J_k|X^n_{t+1},Z^n) \nonumber \\
    &~~~~+I(X_{t};J_{k'}|X^n_{t+1}) + I(X_t;Y^{t-1}_{k'}|J_k,J_{k'},X^n_{t+1}) - I(X_t;J_k|J_{k'},Y^{t-1}_{k'},X^n_{t+1})\} \label{lemma-ll}
\end{align}
 \lipsum[1][0]
    \par\noindent\rule{\textwidth}{0.5pt}
 \lipsum[1][0]
 \vspace{-5mm}
\end{figure*}
where (a) holds due to the Markov chains $Y_{k,t}-(J_k,Y^{t-1}_k,Z^n)-X^{t-1}$,~{$(Y_{k,t},Y_{k',t})-(J_{k'},Y^{t-1}_{k},Y^{t-1}_{k'})-X^{t-1}$} and $(Y_{k,t},Y_{k',t})-(J_k,J_{k'},Y^{t-1}_k,Y^{t-1}_{k'})-X^{t-1}$, (b) follows from the Csisz\'ar sum identity \cite[Ch. 2]{GK}, and (c) is due to reordering {of} the terms.

For the first summation in {\eqref{lemma-ll}}, we have
\begin{align}
    &\sum_{t=1}^n\{-H(X_{t}|Z_t) + H(X_{t}|J_k,Y^{t-1}_k,X^n_{t+1},Z^n) \nonumber \\
    &~~~~- H(X_t) + H(X_{t}|J_{k'},Y^{t-1}_k,Y^{t-1}_{k'},X^n_{t+1}) \nonumber \\
    &~~~~+ H(X_t|J_{k'},Y^{t-1}_{k'},X^n_{t+1}) \nonumber \\
    &~~~~- H(X_t|J_k,J_{k'},Y^{t-1}_k,Y^{t-1}_{k'},X^n_{t+1})\} \nonumber \displaybreak[0]\\
    &\overset{\rm (a)}= \sum_{t=1}^n\{-H(X_{t}|Z_t)  + H(X_{t}|J_k,Y^{t-1}_k,Z_t,Q_t)\nonumber \\
    &~~~~- H(X_t) + H(X_{t}|J_{k'},Y^{t-1}_{k'},Q_t) \nonumber \\
    &~~~~+ H(X_t|J_{k'},Y^{t-1}_{k'},Q_t) \nonumber \\ 
    &~~~~- H(X_t|J_k,J_{k'},Y^{t-1}_k,Y^{t-1}_{k'},Q_t)\} \nonumber \displaybreak[0]\\
    &= \sum_{t=1}^n\{-H(X_{t}|Z_t)  + H(X_{t}|V_{k,t},Q_t,Z_t) - H(X_t) \nonumber \\
    &~~~~+ H(X_t|V_{k',t},Q_t)+H(X_t|V_{k',t},Q_t) \nonumber \\
    &~~~~- H(X_t|V_{k,t},V_{k',t},Q_t)\}\nonumber \displaybreak[0]\\
    &= \sum_{t=1}^n\{-I(X_{t};V_{k,t}|Z_t,Q_t) - I(X_t;V_{k',t}|Q_t) \nonumber \\
    &~~~~+ I(X_t;V_{k,t}|V_{k',t},Q_t)\} \nonumber \displaybreak[0]\\
    &= \sum_{t=1}^n\{I(Z_{t};V_{k,t}|Q_t) - I(X_t;V_{k',t}|Q_t) - I(V_{k,t};V_{k',t}|Q_t)\} \nonumber \displaybreak[0]\\
    &= \sum_{t=1}^n\{I(Z_{t};V_{k,t}|Q_t) - I(X_t;V_{k',t}|Q_t) - \xi'_t\}, \label{aa-b}
\end{align}
where (a) holds due to the following Markov chains:
\begin{align}
    &X_{t}-(J_k,Y^{t-1}_k,Z^n,X^n_{t+1})-X^{t-1}, \\
    &X_{t}-(J_{k'},Y^{t-1}_{k'},X^n_{t+1})-Y^{t-1}_k, \\
    &X_{t}-(J_{k},J_{k'},Y^{t-1}_{k},Y^{t-1}_{k'},X^n_{t+1})-X^{t-1}. \label{73-73}
\end{align}
Note that as a particular case of \eqref{73-73}, we also have $X_{t}-(J_{k'},Y^{t-1}_{k'},X^n_{t+1})-X^{t-1}$.

For the second summation in {\eqref{lemma-ll}}, it follows that
\begin{align}
    &\sum_{t=1}^n\{I(X_t;J_k|X^n_{t+1},Z^n) + I(X_t;J_{k'}|X^n_{t+1}) \nonumber \\
    &~~~~+ I(X_t;Y^{t-1}_{k'}|J_k,J_{k'},X^n_{t+1}) \nonumber \\
    &~~~~- I(X_t;J_k|J_{k'},Y^{t-1}_{k'},X^n_{t+1})\} \nonumber \displaybreak[0] \\
    &=\sum_{t=1}^n\{I(X_t;J_k|X^n_{t+1},Z^n) + I(X_t;J_{k'}|X^n_{t+1}) \nonumber \\
    &~~~~+ H(X_t|J_k,J_{k'},X^n_{t+1}) - H(X_t|J_{k'},Y^{t-1}_{k'},X^n_{t+1})\} \nonumber \displaybreak[0] \\
    &=\sum_{t=1}^n\{I(X_t;J_k|X^n_{t+1},Z^n) + I(X_t;J_{k'}|X^n_{t+1}) \nonumber \\
    &~~~~+ H(X_t|J_k,J_{k'},X^n_{t+1}) - H(X_t|V_{k',t},Q_{t})\} \nonumber \displaybreak[0] \\
    &=\sum_{t=1}^n\{-I(X_t;Z_t) + I(X_t;Z^n|J_k,X^n_{t+1}) \nonumber \\
    &~~~~+ I(X_t;V_{k',t}|Q_{t})\} + I(X^n;J_k) - I(X^n;J_k|J_{k'}) \nonumber \displaybreak[0] \\
    &\overset{\rm (a)}\ge \sum_{t=1}^n\{-I(X_t;Z_t) + I(X_t;Z_t|J_k,X^n_{t+1}) \nonumber \\
    &~~~~+ I(X_t;V_{k',t}|Q_{t})\} + I(J_k;J_k') \nonumber \displaybreak[0] \\
    &\overset{\rm (b)}= \sum_{t=1}^n\{-H(Z_t) + H(Z_t|J_k,X^n_{t+1}) \nonumber \\
    &~~~~ + I(X_t;V_{k',t}|Q_{t})\} + I(J_k;J_k') \nonumber \displaybreak[0] \\
    &\overset{\rm (c)}\ge \sum_{t=1}^n\{-H(Z_t) + H(Z_t|J_k,Y^{t-1}_k,Q_t) \nonumber \\
    &~~~~+ I(X_t;V_{k',t}|Q_{t})\} + I(J_k;J_k') \nonumber \\
    &= \sum_{t=1}^n\{-I(Z_t;V_{k,t}|Q_t) + I(X_t;V_{k',t}|Q_{t})\} + I(J_k;J_k'), \label{aa-c}
\end{align}
where (a) is due to the Markov chain $J_k-X^n-J_{k'}$, (b) is due to the Markov chain $Z_t-X_t-(J_k,X^n_{t+1})$, and (c) follows since conditioning reduces entropy. Substituting \eqref{aa-b} and \eqref{aa-c} into \eqref{lemma-ll}, it follows that
\begin{align}
    F \ge -\sum_{t=1}^n\xi'_t + I(J_k;J_{k'}). \label{fijk}
\end{align}
Lastly, plugging \eqref{fijk} into \eqref{aa-a}, Lemma \ref{lemma-ll-eve} is proved.
\end{proof}

\medskip
For the choice of $V_{k,t}, U_{k,t}$ in \eqref{vkukqt}, similar to \eqref{lemma-yjxn}--\eqref{dhxu1u2q}, we also have
\begin{align}
    &H(J_k|X^n) =\sum_{t=1}^nI(Y_{k,t};U_{k,t}|X_t,Q_t), \label{lemma-yjxn-eve} \displaybreak[0] \\
    &n(\sum_{k \in \mathcal{S}}R_k +2\delta) \ge H(J_{\mathcal{S}}) \ge \sum_{t=1}^n\left\{H(X_t|U_{\mathcal{S}^c,t},Q_t)\right\} \nonumber \displaybreak[0] \\
    &~~~~+ \sum_{k \in \mathcal{S}}\Big\{\sum_{t=1}^nI(Y_{k,t};{U_{k,t}}|X_t,Q_t)\Big\}-n(D + \delta), \label{hjsc1111-eve} \displaybreak[0] \\
    &{n(D+\delta) \ge \sum_{t=1}^nH(X_t|U_{1,t},U_{2,t},Q_t)}.
\end{align}

The analyses of the privacy-leakage rates, their sum rate, and the mixed sum rate are verified as follows: 

\medskip
\noindent{\em {\em i)} Analysis of the Privacy-Leakage Rate}: From the same steps to derive \eqref{lk-delta}, we have
\begin{align}
    n&(L_k + \delta) \ge H(J_{k})-H(J_k|X^n)-I(Z^n;J_k) \nonumber \displaybreak[0]\\
    &\overset{\rm (a)}\ge \sum_{t=1}^n\{H(X_t|U_{k',t},Q_t) - (D+\delta)\} - I(Z^n;J_k) \nonumber \\
    &\overset{\rm (b)} \ge \sum_{t=1}^n\{H(X_t|U_{k',t},Q_t) +I(V_{k,t};U_{k',t}|Q_t) \nonumber \displaybreak[0]\\
    &~~~~ -I(Z_t;V_{k,t}|Q_t) - \xi'_{t}\}-n(D+\delta),
\end{align}
where (a) is due to \eqref{lemma-yjxn-eve} and \eqref{hjsc1111-eve} with $\mathcal{S} =\{k\}$ and $\mathcal{S}^c =\{k'\}$, and (b) follows since Lemma \ref{lemma-ll-eve} applies.

\medskip
\noindent{\em {\em ii)} Analysis of the Sum Rate}: Again from \eqref{tau-1}, it follows that
\begin{align}
    &n(L_1 + L_2 + 2\delta) \nonumber \\
    &\ge I(X^n;J_1) + I(X^n;J_2) - I(Z^n;J_1)-I(Z^n;J_2) \nonumber \displaybreak[0] \\
    &= H(J_1) + H(J_2) - H(J_1|X^n) - H(J_2|X^n) \nonumber \\
    &~~~~- I(Z^n;J_1)-I(Z^n;J_2) \nonumber \displaybreak[0] \\
    &\ge H(J_1,J_2) - H(J_1|X^n) - H(J_2|X^n) \nonumber \\
    &~~~~- I(Z^n;J_1)-I(Z^n;J_2) \nonumber \displaybreak[0] \\
    &\overset{\rm (a)}\ge \sum_{t=1}^n\{H(X_t) - (D+\delta)\} - I(Z^n;J_1)-I(Z^n;J_2) \nonumber \displaybreak[0] \\
    &\overset{\rm (b)}\ge \sum_{t=1}^n\{H(X_t) + I(V_{1,t};U_{2,t}|Q_t) - I(Z_t;V_{1,t}|Q_t) \nonumber \\
    &~~~~+ I(V_{2,t};U_{1,t}|Q_t)-I(Z_t;V_{2,t}|Q_t) - 2\xi'_{t}\} -n (D+\delta)\nonumber \\
    &\ge \sum_{t=1}^n\{H(X_t) + I(V_{1,t};V_{2,t}|Q_t)-I(Z_t;V_{1,t}|Q_t) \nonumber \\
    &~~~~ - I(Z_t;V_{2,t}|Q_t) - \xi'_{t}\}-n(D+\delta),
    \label{l1l2-llg-1}
\end{align}
where (a) follows from \eqref{lemma-yjxn-eve} and \eqref{hjsc1111-eve} for $\mathcal{S} =\{1,2\}$ and (b) is due to Lemma \ref{lemma-ll-eve}.

\medskip
\noindent{\em {\em iii)} Analysis of the Mixed Sum Rate}: As in \eqref{rkdlk},
\begin{align}
    &n(R_{k'} + L_k + 2\delta) \ge H(J_{k'},J_k) - H(J_k|X^n) - I(Z^n;J_k) \nonumber \displaybreak[0]\\
    &\overset{\rm (a)} \ge \sum_{t=1}^n\{I(Y_{k',t};U_{k',t}|X_t,Q_t) + H(X_t) - (D+\delta) \nonumber \\ &~~- I(Z_t;J_k|Z^{t-1})\}\nonumber \displaybreak[0]\\
    &\ge \sum_{t=1}^n\{I(Y_{k',t};U_{k',t}|X_t,Q_t) + H(X_t) - (D+\delta) \nonumber \\
    &~~-~~ I(Z_t;J_k,Y^{t-1}_k,Q_t,Z^{t-1})\}\nonumber \displaybreak[0]\\
    &\overset{\rm (b)}= \sum_{t=1}^n\{I(Y_{k',t};U_{k',t}|X_t,Q_t) + H(X_t) - (D+\delta)\nonumber \\
    &~~~~-I(Z_t;V_{k,t}|Q_t)\},
\end{align}
where (a) follows from {\eqref{lemma-yjxn-eve} and} \eqref{hjsc1111-eve} for $\mathcal{S} =\{k,k'\}$ and $\mathcal{S}^c = \emptyset$ and (b) is due to the i.i.d. property of $(X^n,Z^n)$.

The proof is wrapped up by applying the standard single-letterization argument and the support lemma together with the Fenchel-Eggleston-Carath\'eodory theorem \cite{GK} to bound the cardinalities of $\mathcal{V}_{k}, \mathcal{U}_{k}$, and $\mathcal{Q}$. Finally, the converse proof is completed by letting $n \rightarrow \infty$ and $\delta \downarrow 0$.
\qed
\section{A Counterexample}
In this appendix, we show that the inner and outer bounds on the rate-distortion-equivocation region under the log-loss do not coincide. From \cite[Equations (84)--(92)]{naghibis2015}, the inner bound is given by the closure of the set of all tuples $(R_1,R_2,\Delta_1,\Delta_2,D) \in \mathbb{R}^5_+$, where $\Delta_1$ and $\Delta_2$ denote the equivocation rates as in \cite{naghibis2015}, satisfying
\begin{align}
    R_1 &\ge I(Y_1;\Tilde{U}_1|\Tilde{U}_2,Q),~~~~R_2 \ge I(Y_2;\Tilde{U}_2|\Tilde{U}_1,Q), \nonumber \\
    R_1 + R_2 &\ge I(Y_1,Y_2;\Tilde{U}_1,\Tilde{U}_2|Q), \nonumber \\
    \Delta_1 &\le H(X) - I(X;\Tilde{U}_1|\Tilde{U}_2,Q), \nonumber \\
    \Delta_2 &\le H(X)-I(X;\Tilde{U}_2|\Tilde{U}_1,Q), \nonumber \\
    \Delta_1 + \Delta_2 &\le H(X) + H(X|\Tilde{U}_1,\Tilde{U}_2,Q), \nonumber \\
    \Delta_2 - R_1 &\le H(X)- I(Y_1,X;\Tilde{U}_1,\Tilde{U}_2|Q), \nonumber \\
    \Delta_1-R_2 &\le H(X)-I(X,Y_2;\Tilde{U}_1,\Tilde{U}_2|Q), \nonumber \\
    D &\ge H(X|\Tilde{U}_1,\Tilde{U}_2,Q) \label{A56}
\end{align}
for a joint distribution
\begin{align}
    P_{X}(x)P_{Q}(q)\prod_{{k}=1}^2P_{Y_k|X}(y_k|x)P_{\Tilde{U}_k|Y_kQ}(\Tilde{u}_k|y_k,q) \label{jd-hatu}
\end{align}
with $|\Tilde{\mathcal{U}}_k| \le |\mathcal{Y}_k| + 3$ and $|Q| \le 6$.

By a similar argument shown in Appendix \ref{apd-B-A} with defining $\Tilde{U}_{1,t} = (J_1,Y^{t-1}_1)$, $\Tilde{U}_{2,t} = (J_2,Y^{t-1}_2)$, and $Q_t = X^{n \backslash t}$, the outer bound on the rate-distortion-equivocation region is characterized by the closure of the set of all tuples $(R_1,R_2,\Delta_1,\Delta_2,D) \in \mathbb{R}^5_+$ satisfying
\begin{align}
    R_1 &\ge I(Y_1;\Tilde{U}_1|X,Q)+H(X|\Tilde{U}_2,Q)-D, \nonumber \\
    R_2 &\ge I(Y_2;\Tilde{U}_2|X,Q)+H(X|\Tilde{U}_1,Q)-D, \nonumber \\
    R_1 + R_2 &\ge I(Y_1;\Tilde{U}_1|X,Q)+I(Y_2;\Tilde{U}_2|X,Q) + H(X)-D, \nonumber \\
    \Delta_1 &\le H(X)-H(X|\Tilde{U}_2,Q)+D, \nonumber \\
    \Delta_2 &\le H(X)- H(X|\Tilde{U}_1,Q)+D, \nonumber \\
    \Delta_1 + \Delta_2 &\le H(X) + D, \nonumber \\
    \Delta_2-R_1 &\le - I(Y_1;\Tilde{U}_1|X,Q)+D, \nonumber \\
    \Delta_1 - R_2 &\le - I(Y_2;\Tilde{U}_2|X,Q) + D, \nonumber \\
    D &\ge H(X|\Tilde{U}_1,\Tilde{U}_2,Q) \label{A57}
\end{align}
for a joint distribution defined in \eqref{jd-hatu} with $|\Tilde{\mathcal{U}}_k| \le |\mathcal{Y}_k| + 2$ and $|Q| \le 6$.

\medskip
Suppose that the rate-distortion-equivocation region is tight, that is, the inner bound defined in \eqref{A56} matches the outer bound defined in \eqref{A57}.
However, a counterexample can be constructed to show otherwise. Consider the extreme point where $D=I(Y_1;\Tilde{U}_1|X,Q)+H(X|\Tilde{U}_2,Q)$, which represents the minimum distortion at $R_1 = 0$. For this case, the right-hand side of the constraint for $\Delta_1$ in \eqref{A57} becomes
$\Delta_1 \le H(X)-H(X|\Tilde{U}_2,Q)+D= H(X) + I(Y_1;\Tilde{U}_1|X,Q)$, which
may exceed $H(X)$. This quantity is the maximum achievable value for the constraint $\Delta_1$ in the inner bound \eqref{A56}, particularly when both $\Tilde{U}_1$ and $\Tilde{U}_2$ are constant. Therefore, there may exist some rate tuples $(R_1,R_2,\Delta_1,\Delta_2,D)$ in the outer bound that are not contained in the inner bound.
\qed \label{appendixE}
\section{Region Equivalence}
In this appendix, we show that the inner bound $\mathcal{R}^{\rm L}_{\rm in}$ in Proposition \ref{th-log-loss-in} matches the outer bound $\mathcal{R}^{\rm L}_{\rm out}$ in Proposition \ref{th3}. We prove this corollary without caring about the cardinality bounds on the auxiliary random variables $U_1$, $U_2$, and the time-sharing variable $Q$ since they are obtainable by independent arguments.

\begin{figure*}[t]
\begin{align}
    P_1 &= (0,0,0,0,I(Y_1;U_1|X,Q)+I(Y_2;U_2|X,Q)+ H(X)), \nonumber \\
    P_2 &= (0,I(Y_2;U_2|X,Q),0,0,I(Y_1;U_1|X,Q)+H(X)), \nonumber \\
    P_3 &= (I(Y_1;U_1|X,Q),0,0,0,I(Y_2;U_2|X,Q)+H(X)), \nonumber \\
    P_4 &= (I(Y_1;U_1|X,Q),I(Y_2;U_2|X,Q),0,0,H(X)), \nonumber \\
    P_5 &= (0,I(Y_2;U_2|Q),0,I(X;U_2|Q),I(Y_1;U_1|X,Q)+H(X|U_2,Q)),\nonumber \\
    P_6 &= (I(Y_1;U_1|Q),0,I(X;U_1|Q),0,I(Y_2;U_2|X,Q) +H(X|U_1,Q)), \nonumber \\
    P_7 &= (I(Y_1;U_1|X,Q),I(Y_2;U_2|Q),0,I(X;U_2|Q), H(X|U_2,Q)), \nonumber \\
    P_8 &= (I(Y_1;U_1|Q),I(Y_2;U_2|X,Q),I(X;U_1|Q),0,H(X|U_1,Q)), \nonumber \\
    P_9 &= (I(Y_1;U_1|Q),I(Y_2;U_2|U_1,Q),I(X;U_1|Q),I(X;U_2|U_1,Q),H(X|U_1,U_2,Q)), \nonumber \\
    P_{10} &= (I(Y_1;U_1|U_2,Q),I(Y_2;U_2,Q),I(X;U_1|U_2,Q), I(X;U_2|Q), H(X|U_1,U_2,Q)), \label{p1-p10}
\end{align}
 \lipsum[1][0]
    \par\noindent\rule{\textwidth}{0.5pt}
 \lipsum[1][0]
 \vspace{-5mm}
\end{figure*}

It is sufficient to show the relationship $\mathcal{R}^{\rm L}_{\rm in} \supseteq \mathcal{R}^{\rm L}_{\rm out}$ since we have already demonstrated in Appendix \ref{appendixB} that every achievable tuple with respect to Definition \ref{defsystem} is contained in the region $\mathcal{R}^{\rm L}_{\rm out}$. By the same approach in \cite{courtade2014}, for $j \in [1:10]$, we say that a point ${P_j} = (R_1^{(j)},R_2^{(j)},L_1^{(j)},L_2^{(j)},D^{(j)}) \in \mathcal{R}^{\rm L}_{\rm out}$ is dominated by a point $\mathcal{R}^{\rm L}_{\rm in}$ if there exists a rate tuple $(R_1,R_2,L_1,L_2,D) \in \mathcal{R}^{\rm L}_{\rm in}$ for which $R_1 \le R_1^{(j)}$, $R_2 \le R_2^{(j)}$, $L_1 \le L_1^{(j)}$, $L_2 \le L_2^{(j)}$, and $D \le D^{(j)}$. Fixing $P_Q$, $P_{U_1|Y_1Q}$, and $P_{U_2|Y_2Q}$, the extreme points of the polytope defined by the constraints in $\mathcal{R}^{\rm L}_{\rm out}$ are listed in \eqref{p1-p10} at the top of this page, where  the points $P_1,\cdots,P_8$ correspond to the endpoints when the constraints $R_1 + R_2$, $R_1 + L_2$, $R_2 + L_1$, $L_1 + L_2$, $R_1$, $R_2$, $L_1$, and $L_2$ in $\mathcal{R}^{\rm L}_{\rm out}$ become zero, respectively, and $P_9$ and $P_{10}$ are attained when the distortion constraint is optimal.

Note that the points $P_9$ and $P_{10}$ lie in the region $\mathcal{R}^{\rm L}_{\rm in}$. Now observe that the point $(0,0,0,0,H(X))$ is in $\mathcal{R}^{\rm L}_{\rm in}$, obtained by setting $U_1$ and $U_2$ to be constant, and this point dominates the points $P_1$, $P_2$, $P_3$, and $P_4$. In case of setting only $U_1$ to be constant, the point $$(0,I(Y_2;U_2|Q),0,I(X;U_2|Q),H(X|U_2,Q))$$ is in $\mathcal{R}^{\rm L}_{\rm in}$ {and} dominates $P_5$ and $P_7$. Moreover, by setting $U_2$ to be constant, we have that the point $$(I(Y_1;U_1|Q),0,I(X;U_1|Q),0,H(X|U_1,Q))$$ {is included} in $\mathcal{R}^{\rm L}_{\rm in}$, {which} dominates $P_6$ and $P_8$. Since $\mathcal{R}^{\rm L}_{\rm out}$ is the convex hull of these extreme points $P_1,\cdots,P_{10}$, we have $\mathcal{R}^{\rm L}_{\rm in} \supseteq \mathcal{R}^{\rm L}_{\rm out}$.
\qed

\end{document}